\pdfoutput=1
\documentclass[osajnl,twocolumn,showpacs,superscriptaddress,10pt]{revtex4-1} 
\usepackage{amsmath,amssymb,graphicx}
\begin{document}

\title{Energy Redistribution Signatures in Transmission Microscopy of Rayleigh- and Mie-particles}

\author{Markus Selmke}\email{Corresponding author: selmke@rz.uni-leipzig.de}
\author{Frank Cichos}
\affiliation{Universit\"at Leipzig, Experimental physics I, molecular nanophotonics, Linn\'estr.\ 5, 04103 Leipzig, Germany}

\begin{abstract}
The transmission characteristics of a spherical (possibly multilayered) particle of arbitrary size under focused illumination are discussed within the generalized Lorenz-Mie theory. Expressions generalizing the total extinction and scattering cross-sections to their fractional counterparts are presented which allow for a convenient modeling of transmission signals, both on-axis and off-axis. The strong dependence of the signal on the collection angle and the complex polarizability are readily included in this minimal, yet accurate model. The precise signature of the energy redistribution and absorption are found for particles of arbitrary complex-valued polarizability. For perfect dielectrics, a transition from sensitive extinction to insensitive scattering signals is observed and quantified for apertures including angles larger than twice the beam's angle of divergence. Implications for positioning, temperature control, spectroscopy and optimized extinction measurements are discussed.
\end{abstract}

\maketitle 

\section{Introduction}
\begin{figure}[bth]
\centerline{\includegraphics [width=\columnwidth]{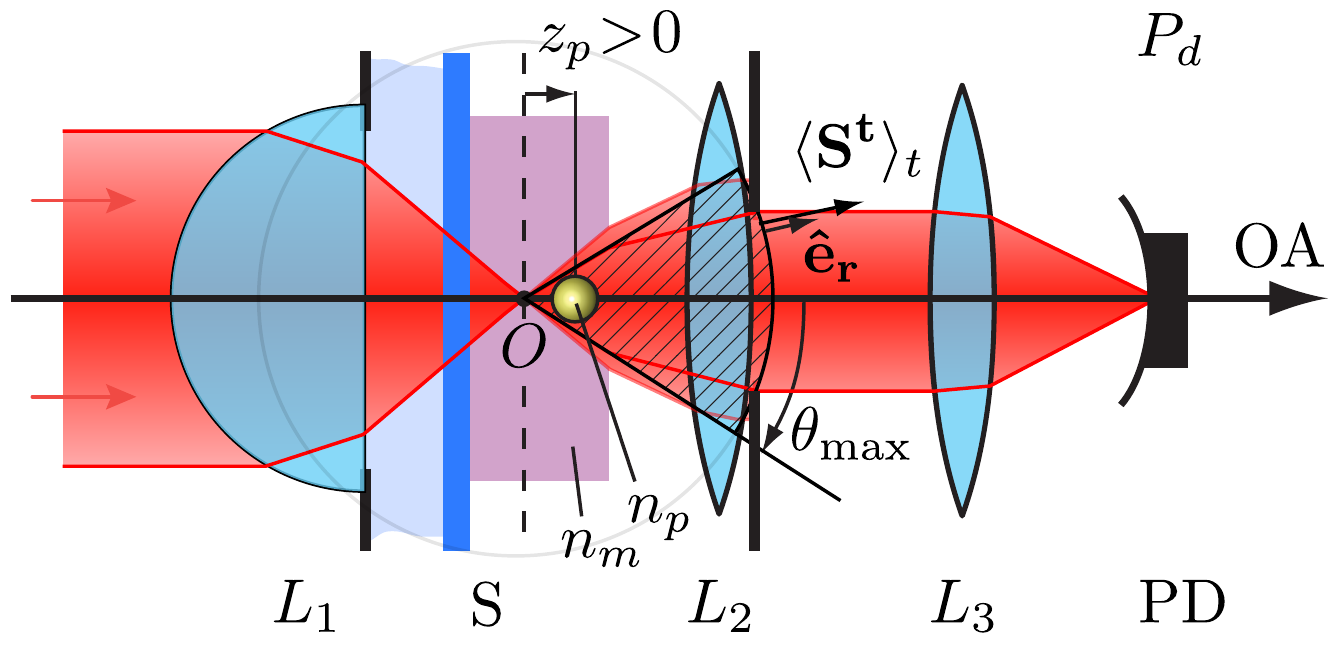}}
\caption{Schematic of a transmission microscope setup. The two microscope objective lenses $L_1$ and $L_2$ form an afocal system with a focal plane inside the sample S intersecting the optical axis OA at the origin $O$. The transmitted and scattered beam is collected by $L_2$ up to an angle $\theta_{\rm max}$ (dashed area, determined by the numerical aperture ${\rm NA}_d=n_m \sin\left(\theta_{\rm max}\right)$) and imaged via lens $L_3$ onto a photodetector measuring $P_d$.}\label{Fig:Setup}
\end{figure}

The optical transmission signal of spherical particles under focused coherent illumination is an informative and conveniently measurable quantity in transmission microscopy setups, see Fig.\ \ref{Fig:Setup}. The power of the transmitted electromagnetic beam is detected in the far-field using a single photodetector and can be understood as the self-interference of the incidence beam with the scattered field \cite{Taubenblatt1991,HwangMoerner2007}. Using the spatial modulation spectroscopy (SMS) technique \cite{Arbouet2004,Billaud2008,Lerme2008,Muskens2008}, the method is very sensitive and well-suited for particle extinction spectroscopy. For single molecules the system is better described by a two-level-system or a classical perfect dipole and may exhibit practically no absorption but strong scattering and even perfect reflection \cite{Zumofen2008}. Under certain conditions, metallic nanoparticles exhibit similar characteristics \cite{Mojarad2009,Wennmalm2012}. A transmission microscopy setup may also be used as a photothermal microscope to indirectly detect single absorbing particles via the creation of a thermal lens with the introduction of a second focused pump laser \cite{Berciaud2006,SelmkeACSNano,NanoLensDiff}. 

Such interference microscopy schemes have been modeled in the Rayleigh-limit of small particles as compared to the wavelength of light, i.e.\ $R\ll \lambda$. The plasmonic-relevant case of a complex-valued polarizabilities was first done by Taubenblatt and Batchelder \cite{Taubenblatt1991} and also recently by Hwang and Moerner \cite{HwangMoerner2007}, who considered particles placed on the optical axis only. Especially the latter treatment unveiled the important role of the particle placement within the focal region and revealed the existence of two kinds of shapes, dip-like and dispersive, for the signal. Similar to perfect dielectric particles \cite{Pralle1999}, the transmission signal off resonance must not be minimal if the particle is in-focus and that, in general, an axial signature is obtained showing characteristics of both patterns. However, these approaches take only the experimentally less important case of detection with a low numerical aperture microscope objective, i.e. ${\rm NA}_d=0$, and point-like particles into account. On the other hand, the corresponding high-${\rm NA}_d$ treatments for idealized dipole scatterers and two-level systems \cite{Zumofen2008,Vamivakas2011} cannot be directly transferred to absorbing induced-dipole (Rayleigh) scatterers \cite{Mojarad2009}.
For large particles $R\gtrsim \lambda/20$ or complex-valued polarizabilities the situation is more complicated. A quantitative but laborious description in the framework of vectorial focusing and coherent scattering has been given by Rohrbach and Stelzer \cite{Rohrbach2002} and later for Laguerre-Gaussian beams by T\"or\"ok et al.\ \cite{Torok2007}. Further, the work by Lerm\'e et al. \cite{Lerme2008,Lerme22008} provided analytical expressions for the transmitted powers and angular distributions of transmitted intensities important for SMS in a multipole expansion of the fields. A similarly rigorous approach based on the vectorial diffraction framework was chosen for the treatment of perfect dipoles and nanoparticles under tight focusing in Ref.\ \cite{Mojarad2009}. However, both works do not consider the axial particle placement in the focal region nor discuss the crucial role played by the collection angle.

Within the framework of the generalized Lorenz-Mie theory (GLMT) \cite{Gouesbet1988,Gouesbet2011} a minimal yet accurate description for a particle of arbitrary size positioned arbitrarily within the focal region is accessible and simplifies matters considerably, especially for focused Gaussian beams \cite{LockTightFocusing} and small particles. While it is common to evaluate the total energy fluxes associated with particle scattering in terms of (total) cross-sections, only fractions of these fluxes can be measured in a microscope setup. It is the aim of this paper to supplement the versatile and accurate framework of the GLMT by convenient and compact expressions that allow the computation of such transmission signals commonly encountered in single particle (photothermal) microscopic and spectroscopic investigations. A simplification of the rigorous theory provides in detail the missing link to a simple concept of a driven dipole field interfering with the incident field. The exact signature of a partial energy redistribution is found. Also, In view of the various misconceptions regarding either the amplitudes and phases involved \cite{HwangMoerner2007} or the neglect of the energy redistribution character of the signal \cite{Pralle1999,Berciaud2006}, some clarification of the details of the physical signal origin appear to be in order.

\section{Theoretical Background}

\subsection{Interference of a driven multipole}
The general picture of what happens in a transmission microscope setup is readily illustrated considering a Gaussian beam incident onto a small and electrically polarizable particle. The incident beam acquires a total phase advance of $\phi_G\rightarrow \pi$ as compared to a plane wave in the far field due to the Gouy-effect \cite{HwangMoerner2007}. Half of this value is accumulated up to the focal point, and the other half behind it. Depending in the particle's optical properties, the driven dipole radiates a scattered field with a certain phase lag relative to the local phase of the incidence beam. The result is a standing wave in backwards direction and a total phase-difference in the forward direction which is determined by the particle properties and its position in the beam. The interference can either be constructive or destructive \cite{HwangMoerner2007,Pototschnig2011}, leading to a reduced or enhanced transmission signal, see Fig.\ \ref{Fig:interference}. If a nanoparticle represents the oscillator, a resonant excitation will further lead to a net energy-uptake, i.e.\ absorption, further accounting for the reduced transmission. Considering the steady-state Poynting theorem $\nabla \cdot \mathbf{S^t}+\mathbf{E^i}\cdot \mathbf{J}=0$, the absorbed energy accounts for the work done by the field on the driven charge carriers \cite{Malvaldi2009}, which constitute a current density $\mathbf{J}$ against the instantaneous direction of the field for finite phase lags $\phi_{\rm sca}$. For a perfect dielectric the relative phase of the scattered field is zero and the dipole oscillates in-phase and loss-less. If no absorption takes place, the situation simply represents an energy redistribution via interference.

For larger particles the concept of a polarizability is no longer sufficient as higher order multipoles then contribute. However, the general situation is similar in a way that both absorption and energy-redistribution determine the transmission of an illuminating focused beam. In addition, scattering becomes important and in the case of focused illumination it even accounts for near field shadow effects \cite{Locke1995}. The energy-redistribution for metallic particles then transitions to the extreme of a backwards reflection according to the Fresnel coefficients and a near-perfect cancellation in the forward direction, both accounted for by the multipolar scattered field, see Fig.\ \ref{Fig:interference}b). 

\subsection{Theory of transmission signals in the GLMT\label{sec:Mie}}
\begin{figure}[t]
\centerline{\includegraphics [width=\columnwidth]{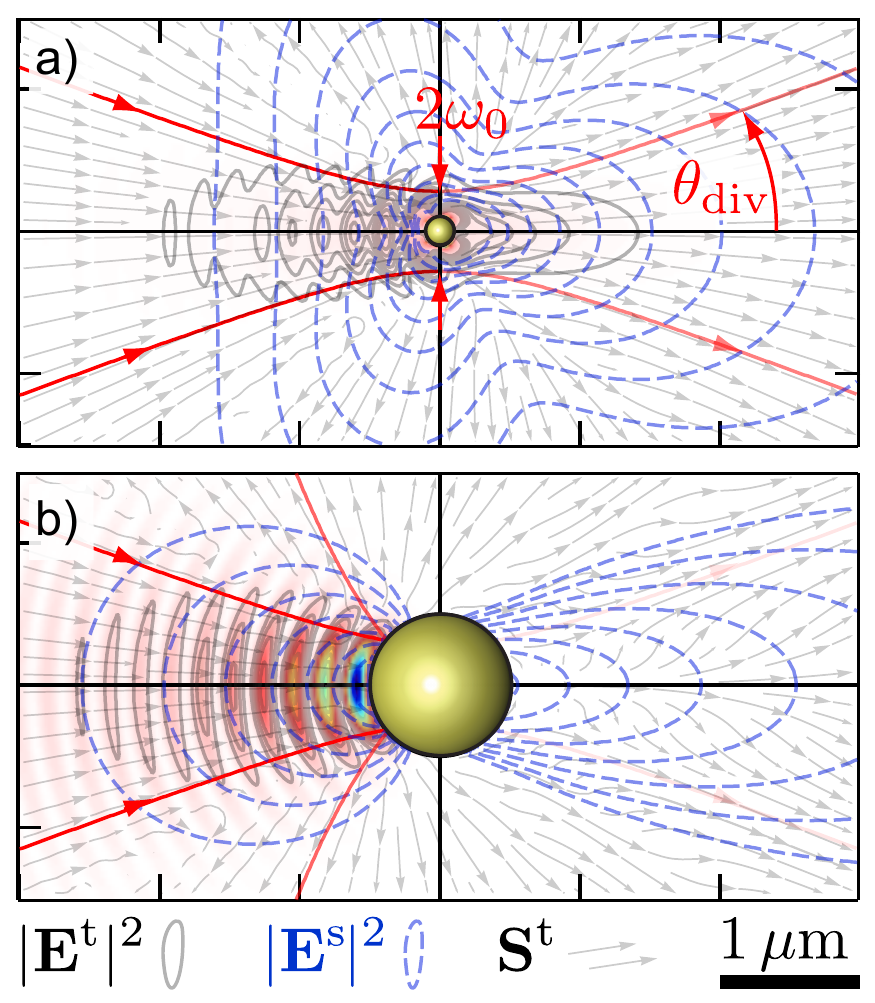}}
\caption{Scattering and interference of a focused Gaussian beam by a AuNP with $R=100\,\rm nm$ in a medium with $n_m=1.46$. The scattered field $|\mathbf{E^s}|^2$ (blue dashed contours) interferes with the incident beam (red / green) to a total field (black contours $|\mathbf{E^t}|^2$ and Poynting vectors $\mathbf{S^t}$). \textbf{a)} Non-resonant illumination with a beam at $\lambda=635\,\rm nm$ and beam-waist $\omega_0=281\rm nm$. The quasi static polarizability is $\alpha_{635}= \left(2.28 - 1.60i\right)\times 10^7\,{\rm nm}^3$. \textbf{b)} Same as a), but for $R=500\,\rm nm > \omega_0$. The incident beam is almost entirely reflected forming a standing wave. The scattered field acts as a compensating field in the forward direction where $\mathbf{E^s}\approx -\mathbf{E^i}$.}\label{Fig:interference} 
\end{figure}

The generalized Lorenz-Mie theory (GLMT) \cite {Gouesbet1988,Gouesbet2011} and its extension to multilayered spheres describes the exact solution to the Maxwell equations for a scattering process with a shaped time-harmonic beam. As the theory solves for the scattered electromagnetic fields $\mathbf{E^s}$ and $\mathbf{H^s}$, the resulting total field $\mathbf{E^t}=\mathbf{E^i}+\mathbf{E^s}$ may be used to compute associated fluxes of electromagnetic energy in a given direction, see Fig.\ \ref{Fig:interference}. This allows a precise modelling of what has been introduced only qualitatively above.

The radial component of the total field's Poynting vector $\mathbf{S^t}=\mathbf{E^t}\times\mathbf{H^t}\equiv \mathbf{S^{\rm i}} +  \mathbf{S^{\rm s}} +  \mathbf{S^{\rm ext}}$ describes this energy flux, and may be evaluated in the far-field as a time-average $\langle \rangle_t$ to find the power contained within a polar angle $\theta_{\rm max}$:
\begin{equation}
P_d=\lim_{r\rightarrow \infty }\int_{0}^{2\pi}\!\!\int_{0}^{\theta_{\rm max}} r^2 \,\mathbf{\hat{e}_r} \cdot \langle\mathbf{S^t}\left(r,\phi,\theta\right)\rangle_t\, \mathrm{d}\Omega.\label{eq:Pd}
\end{equation}
In the language of Mie theory, the total power is typically decomposed into three constituents according to $P_d=P_{\rm inc}+P_{\rm sca}+P_{\rm ext}=I_0\left[\sigma_{\rm inc}+\sigma_{\rm sca}-\sigma_{\rm ext}\right]$, i.e.\ incidence, scattering and extinction, respectively. The latter term represents the interference of the incidence and the scattered field. The measurable quantity of interest is the relative signal compared to the background, i.e. $P_d=P_{\rm inc}$ when no particle is obstructing the beam,
\begin{equation}
\frac{P_d}{P_{\rm inc}} - 1 = \frac{\Delta P_d}{P_{\rm inc}} = \frac{\sigma_{\rm sca}\left(\theta_{\rm max}\right) - \sigma_{\rm ext}\left(\theta_{\rm max}\right)}{\sigma_{\rm inc}\left(\theta_{\rm max}\right)}.\label{eq:PdRel}
\end{equation}
The incident beam field is represented as a series of eigenfunctions satisfying the Maxwell equations and is specified by a set of expansion coefficients, which in case of an axisymmetric beam are the single-indexed beam shape coefficients (BSCs) $g_n$. For plane-waves they are unity, i.e.\ $g_n=1$. The most convenient assumption for a focused illumination is the Gaussian beam,
\begin{align}
\mathbf{E^i}\left(\rho,z\right)&\approx-\mathbf{\hat{e}_x} \frac{E_0 \exp\left(-ikz\right)}{1-i z/z_R}\exp\left(\frac{-\rho^2/\omega_0^2}{1-iz/z_R}\right)\label{eq:IncField},
\end{align}
with a beam waist $\omega_0$, Rayleigh-range $z_R=k\omega_0^2/2$ and wavenumber $k=n_m 2\pi/\lambda$ in the embedding medium with index of refraction $n_m$. The Gouy phase $\phi_G=\arctan\left(z_p/z_R\right)$ is contained in the exponential prefactor. Accordingly, the axial intensity profile is described by a Lorenzian, $I = I_0/\left[1+z^2/z_R^2\right]$. The following BSCs well describe the incidence field \cite{Gouesbet2011,LockTightFocusing}:
\begin{equation}
g_n=Q\exp\left(-Q s^2\left[n-1\right]\left[n+2\right]\right)\exp\left(-i k z_p\right)\label{eqn:gn}.
\end{equation} 
Herein, $Q=\left(1-i z_p/z_R\right)^{-1}$ and the beam-confinement factor is defined as $s=\omega_0/\left(2z_R\right)$, which is the ratio of the lateral to the axial extent of the beam focus. This quantity is related to the half-angle of divergence $\theta_{\rm div}=2/k\omega_0=2s$. Actually, these BSCs are based on the first-order Davis beam, i.e.\ the corrected Gaussian beam, but anticipate even higher order terms. The axial coordinate of the particle relative to the beam-waist position has the following sign-convention: $z_p <0$ corresponds to a particle being positioned in the converging part of the focused beam, i.e.\ in front of the beam waist in the direction of beam propagation, see Fig. \ref{Fig:Setup}.
The fractional cross-sections as defined via eq.\ \eqref{eq:Pd} simplify to \cite{SelmkeACSNano,NanoLensDiff}:
\begin{align}
\sigma_{\rm inc}=&\frac{\pi}{k^2}\,\frac{2}{\theta_{\rm div}^2} \left[1 - \exp\left(-2\tan^2\left(\theta_{\rm max}\right)/\theta_{\rm div}^2\right)\right]\!,\label{eq:sigmainc}\\
\sigma_{\rm sca}=&\, \frac{\pi}{k^2}\int_{0}^{\theta_{\rm max}}\left[|S_{1}\!\left(\theta\right)|^2+|S_{2}\!\left(\theta\right)|^2\right]\sin\left(\theta\right) \mathrm{d}\theta\label{eqnScatter},\\
\sigma_{\rm ext}=&\frac{\pi}{k^2}\int_{0}^{\theta_{\rm max}}\!\!\!\!\!\!\mathfrak{R}\left(M^{*}\!\left(\theta\right)\left[S_1\!\left(\theta\right)+S_2\!\left(\theta\right)\right]\right)\sin\left(\theta\right) \mathrm{d}\theta\label{eqnISI}.
\end{align}
Eq.\ \eqref{eq:sigmainc} represents an approximation for a paraxial Gaussian beam and $\theta_{\rm max}\le \pi/2$ only. For an arbitrary axisymmetric beam see Appendix Eq.\ \eqref{eq:AppendixInc}. The total cross-sections $\sigma_{-}^\pi$ are contained in the above expressions in the limit $\theta_{\rm max}=\pi$ and can be expressed by sums over all multipoles $n$ \cite{Gouesbet1988}. In this case, Eq.\ \eqref{eq:Pd} represents an energy balance with the absorbed power $-P_{\rm abs}=-I_0\sigma_{\rm abs}^\pi$ replacing $P_d$.
%
In the previous expressions, the LM scattering functions $S_{1,2}\left(\theta\right)$ and an additionally defined auxiliary function $M\left(\theta\right)$ describing the interference contribution are:
\begin{align}
S_{1}&=\sum_{n=1}^{\infty}N_n\, g_n \left[a_n \Pi_n\left(\theta\right)+b_n \tau_n\left(\theta\right)\right],\label{eq:S1}\\
S_{2}&=\sum_{n=1}^{\infty}N_n\, g_n \left[a_n \tau_n\left(\theta\right)+b_n \Pi_n\left(\theta\right)\right],\label{eq:S2}\\
M\left(\theta\right)&=\sum_{n=1}^{\infty}N_n\,g_n\, \left[\Pi_n\left(\theta\right)+\tau_n\left(\theta\right)\right],\label{eqnM}
\end{align}
%
with $N_n=\left[2n+1\right]/\left[n\left(n+1\right)\right]$. The usual Mie scattering coefficients \cite{BohrenHuffman} $a_n$ and $b_n$ can be substituted by the the outmost layer scattering coefficients for multilayered particles. A public c-code provides these conveniently, see Ref.\ \cite{Pena2009}. The occurring angular functions $\Pi_n\left(\theta\right)$ and $\tau_n\left(\theta\right)$ can be determined recursively and the expressions may be found in the same reference. 

\begin{figure}[tb]
\centerline{\includegraphics [width=\columnwidth]{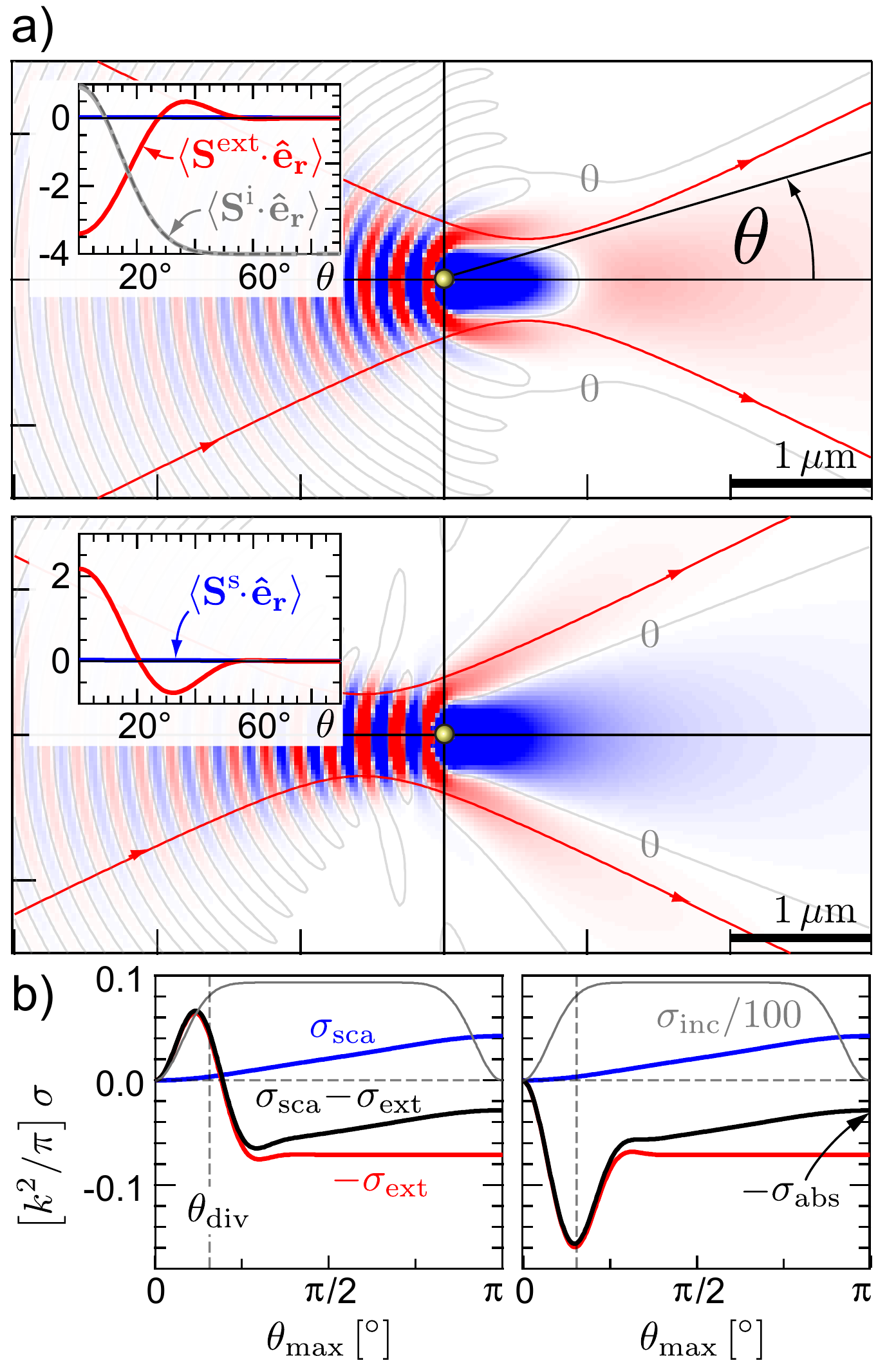}}
\caption{Energy redistribution $|\mathbf{E^t}|^2-|\mathbf{E^{i}}|^2$ ($>0$: blue, $<0$: red) for a $R=30\,\rm nm$ AuNP in a Gaussian beam with $\omega_0=281\,\rm nm$ at $\lambda=635\,\rm nm$, $n_m=1.46$. \textbf{a)} The particle is offset by $z_p=-z_R$ (top), leading to a negative relative transmission $\Delta P_d/P_{\rm inc} < 0$ and $z_p=+z_R$ (bottom), resulting in a positive signal in forward direction. \textbf{b)} Dependence of the fractional cross-sections as a function of the collection angle $\theta_{\rm max}$. The dashed gray line shows the divergence angle $\theta_{\rm div}\approx 28^{\circ}$.}\label{Fig:Redistribution}
\end{figure}

\section{Interference signals in the GLMT}
In view of the physical interpretation of each contribution to the detected power, the extinction term is of special importance as it accounts for the absorption and further embodies the interference causing a spatial energy redistribution of the propagating fields. As already suggested by A.\ Rohrbach et al.\ \cite{Rohrbach2002}, its strong angular dependence on the collection angle $\theta_{\rm max}$ is a consequence of the changing phase relation between the interfering incidence and scattered electric fields in dependence on the polar angle. Using the framework of the GLMT this energy redistribution may be visualized in the near-field as well, see Fig.\ \ref{Fig:Redistribution}a) for the case of a AuNP. The near field approaches the far-field signal distribution already for small distances. In the far field, the resulting fractional cross-sections are shown for two particle offsets in Fig.\ \ref{Fig:Redistribution}b). The change of the total fractional power $\Delta P_d\left(\theta_{\rm max}\right)\propto \sigma_{\rm sca}-\sigma_{\rm ext}$ shows an initial signature of the energy redistribution, i.e.\ interference, in the propagation direction of the incident beam for angles $\theta \lesssim 2\theta_{\rm div}$. Hereafter it is determined by the scattering contribution to eventually saturate at a negative finite value, corresponding to the power absorbed by the NP. For large particles the scattering term dominates and even accounts for a perfect shadow behind an opaque particle completely blocking a focused beam \cite{Locke1995}, i.e.\ for $R\gg \omega_0$, see Fig.\ \ref{Fig:interference}b).

While the integrands of the fractional cross-sections $\sigma_{\rm sca}$ and $\sigma_{\rm ext}$ in \eqref{eqnScatter} and \eqref{eqnISI} may be evaluated to find the angular characteristics, it turns out that the integration may be done analytically. The necessary definite integrals have been encountered in plane-wave Mie theory before \cite{Wiscombe,BabenkoBook}. We here report the resulting expressions when applied to focused beams in the GLMT, see appendix Eqs.\ \eqref{eq:AppendixSigmaExtOnAxis1} - \eqref{eq:AppendixSigmaScaOnAxis}. These rigorous solutions allow a simplification and insight to be gained in the case of small-particle interference. Especially the important interference contribution can thus be studied in its dependence on the collection angle $\theta_{\rm max}$, thereby exceeding the capabilities of a simple dipole model. Further, the expressions provide the means to compute the measurable transmission signals within the framework of the GLMT for an arbitrary sized-scatterer and speed up the calculations dramatically as compared to the numerical procedure reported in our earlier work \cite{SelmkeACSNano,NanoLensDiff}.

\subsection{Small particle approximation (on-axis)\label{sec:SmallParticleApproxAtempt}}
\paragraph{The complex-valued polarizability}
The full solid angle integration of the electromagnetic power fluxes for plane wave illumination yields \cite{BohrenHuffman} the Rayleigh-results for small particles, i.e.\ $\sigma_{\rm sca}^{R}\approx k^4 |\alpha|^2/ 6\pi$ for the scattering cross-section, $\sigma_{\rm ext}^{R}\approx - k\, \mathfrak{I}\left(\alpha\right)$ for the extinction cross-section and consequently $\sigma_{\rm abs}^R \approx \sigma_{\rm ext}^{R} - \sigma_{\rm sca}^{R}$ for the absorption cross-section. They only depend upon the complex-valued electric polarizability $\alpha$ of the particle, i.e.\ the quantity which relates the induced dipole moment to a homogeneous incident field in $\mathbf{p}=\epsilon \alpha \mathbf{E^i}$, with $\epsilon=\epsilon_0 n_m^2$. For a sphere of radius $R$ the Clausius-Mossotti result is:
\begin{equation}
\alpha=|\alpha|\exp\left(i\phi_{\rm sca}\right) \approx 4\pi R^3 \frac{n_p^2-n_m^2}{n_p^2+2n_m^2}\label{eq:alpha},
\end{equation}
which must be corrected by a radiation-reaction term or by relating it to the electric dipolar Mie coefficient $\alpha= 6\pi a_1/i k^3$ \cite{Moroz2010}. The chosen time-dependence $\exp\left(+i\omega t\right)$ dictates the sign of the particle's complex refractive index as $n_p=n - i \kappa$ in Eq.\ \eqref{eq:alpha}. The polarizability determines wether the induced dipole $\mathbf{p}$ oscillates (and thereby radiates) with or without a phase-lag relative to the local polarizing incidence field. For a resonant particle the imaginary part is negative and large in magnitude such that $\phi_{\rm sca}=\arctan\left(\mathfrak{I}\left(\alpha\right)/\mathfrak{R}\left(\alpha\right)\right)\approx - \pi/2$, i.e.\ the phase of the scattered field lags behind the driving. Far from resonance or for a perfect dielectric, the induced dipole radiates in-phase ($\phi_{\rm sca}=0$) and without losses.

\paragraph{The scattering contribution \& absorption}
For a focused beam, the fractional scattering and extinction cross-sections and the total absorption cross-section are of interest. The latter one results from the corresponding energy balance as $\sigma_{\rm abs}^\pi \approx |g_1|^2 \sigma_{\rm abs}^{R}$, i.e.\ the Rayleigh result apart from the prefactor. For a Gaussian beam, this factor is simply proportional to the intensity at the particle's position, i.e.\ $|g_1|^2\propto 1/\left[1+z_p^2/z_R^2\right]$. The fractional scattering cross-section, Eq.\ \eqref{eqnScatter}, simplifies to
\begin{equation}
\sigma_{\rm sca}\approx \frac{|g_1|^2 |\alpha|^2 k^4}{48\pi}\left[4-3\cos\left(\theta_{\rm max}\right)-\cos^3\left(\theta_{\rm max}\right)\right],\label{eq:RAfractionalcrosssections}
\end{equation}
where the series for the scatter functions $S_{1,2}$ have been truncated at the first dipolar contribution corresponding to $n=1$. As usual, it was recognized that the first electric dipolar Mie scatter coefficient $a_1\approx i\alpha k^3/6\pi \propto x^3$, whereas the magnetic dipolar coefficient $b_1$ is already of higher order in the size-parameter $x=kR$. The fractional scattering cross-sections is related to the Rayleigh-result for the scattered power within a forward domain apart from the factor $|g_1|^2$, i.e.\ $\sigma_{\rm sca} =|g_1|^2 \sigma_{\rm sca}^{R}\left(\theta_{\rm max}\right)$.

The previous expression is the expected result for a dipole moment $\mathbf{p}=\epsilon \alpha \mathbf{E^i}\left(z_p\right)$ induced by the incident beam field Eq.\ \eqref{eq:IncField} at the position of the particle. In the far field, the dipole radiates a time-harmonic field of the form $\mathbf{E^p} \rightarrow \left[k^2/4\pi\epsilon\right] \left(\mathbf{\hat{e}_{\mathbf{r}}}\times \mathbf{p}\right)\times \mathbf{\hat{e}_{\mathbf{r}}} \exp\left(-ikr\right)/r$. Indeed, the scattered field in the forward direction in the GLMT can be written as (appendix, Eq.\ \eqref{Efieldsca})
\begin{equation}
\mathbf{E^s}\left(\mathbf{r}\right) \rightarrow -\mathbf{\hat{e}_x} \frac{E_0}{kr}\exp\left(-ikr\right)g_1 \frac{\alpha k^3}{4\pi},\label{eq:ScaField}
\end{equation}
which is the dipole far field as induced by the local Gaussian beam's electric field $\mathbf{E^i}\left(z_p\right) \approx -\mathbf{\hat{e}_x} E_0 g_1\left(z_p\right)$, Eq.\ \eqref{eq:IncField}. Integration of $|\mathbf{E^p}|^2$ over the full azimuthal range yields the GLMT result Eq.\ \eqref{eq:RAfractionalcrosssections}.

\begin{figure}[bt]
\centerline{\includegraphics [width=\columnwidth]{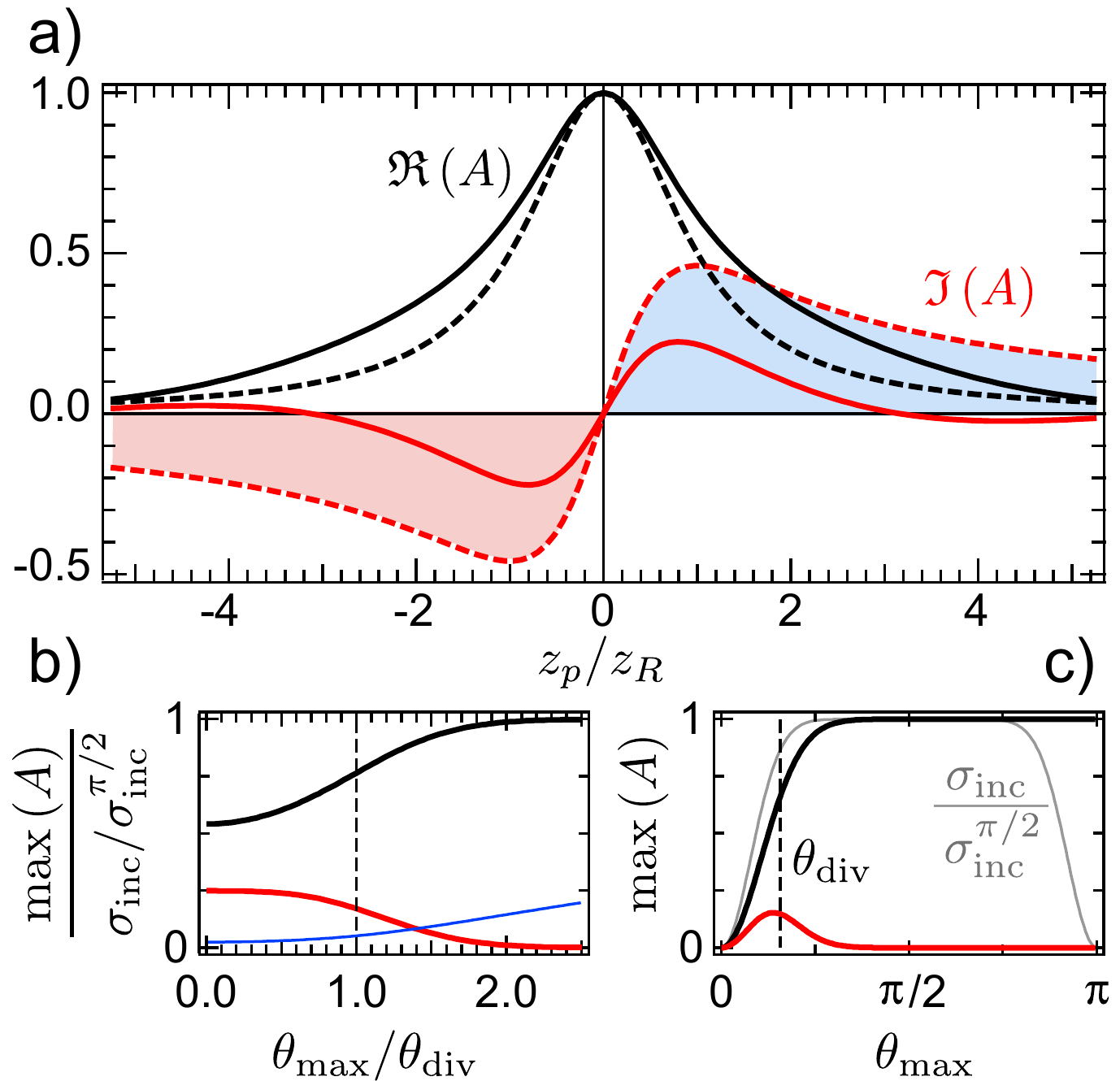}}
\caption{\textbf{a)} Real (black) and imaginary part (red) of the detection function $A\left(z_p\right)$ for low-${\rm NA}_d$ detection ($\theta_{\rm max}=0$, dashed lines) and high-${\rm NA}_d$ detection $\theta_{\rm max}=\theta_{\rm div}$ (solid lines), each normalized to $A\left(0\right)=1$. For $\theta_{\rm max}\gtrsim 2\theta_{\rm div}$ the real part matches the curve for $\theta_{\rm max}=0$ while the imaginary part is zero. \textbf{b)} Evolution of the relative transmission signal, $\left[\Delta P_d/P_{\rm inc}\right]\sigma_{\rm inc}^{\pi/2}/k\alpha$. For comparison, the blue line shows the scattering contribution for $k^3\alpha=1$. \textbf{c)} Plot of the magnitudes ${\rm max}\left(\mathfrak{R}\left(A\right)\right)$ (black) and ${\rm max}\left(\mathfrak{I}\left(A\right)\right)$ (red) of the detection function for different collection angles $\theta_{\rm max}\in \left[0,\pi\right]$.}\label{Fig:Azp}
\end{figure}

\paragraph{The extinction / interference contribution}
Since the first scatter coefficient $a_1\propto x^3$ appears quadratically in the scatter contribution but linear in the extinction term, the interference term usually dominates and therefore determines the shape and magnitude of the relative transmission signal Eq.\ \eqref{eq:PdRel} for small particles, i.e.\ $\Delta P_d/P_{\rm inc}\approx -\sigma_{\rm ext}/\sigma_{\rm inc}$. However, the interference term of a focused beam requires more multipole orders for a proper representation of the incidence field. Therefore, while the scatter functions can still be truncated at the $n=1$-term, one must not do the dame with the sum in $M\left(\theta\right)$, Eq.\ \eqref{eqnM}. The factional extinction cross-section finally becomes $\pi k^{-2}\mathfrak{R}\left(6 a_1 A\right)$, see Appendix. Expressed via the complex-valued polarizability $\alpha$ this amounts to
\begin{align}
\sigma_{\rm ext}\approx& -k \left[\mathfrak{I}\left(\alpha\right)\mathfrak{R}\left(A\right) + \mathfrak{R}\left(\alpha\right)\mathfrak{I}\left(A\right)\right],
\label{eq:RAextapprox}
\end{align}
wherein the first part is seen to equal $\sigma_{\rm abs}^R \mathfrak{R}\left(A\right)$. In the above stated result $A=A\left(z_p\right)$ is a complex-valued amplitude which depends on the particle displacement $z_p$. It is characteristic of the beam used as specified by the BSCs $g_n$ and depends critically on the collection angle. For a Gaussian beam, the function exhibits a dispersive imaginary part and a dip-like real part, see Fig.\ \ref{Fig:Azp}a). The weight of each form comprising the detectable signal is given by the real and imaginary parts of the polarizability. 

\paragraph{Transmission at small angles}
For vanishingly small collection angles $\theta_{\rm max}\approx 0$, the peaks of the dispersive imaginary part are about half in amplitude as compared to the peak-value of the real part. The functional form may then be shown to reduce to (see Appendix)
\begin{align}
\mathfrak{R}\left(A\right) \approx& \frac{\theta_{\rm max}^2}{\theta_{\rm div}^2} \frac{1}{1+z_p^2/z_R^2}, \label{eq:ARe}\\
\mathfrak{I}\left(A\right) \approx& \frac{\theta_{\rm max}^2}{\theta_{\rm div}^2} \frac{z_p/z_R}{1+z_p^2/z_R^2}.\label{eq:AIm}
\end{align}
%



Supplementing the fractional extinction, eq.\ \eqref{eq:RAextapprox}, with the near-forward incidence cross-section $\sigma_{\rm inc} \approx 4\pi \theta_{\rm max}^2/k^2 \theta_{\rm div}^4$ for small angles $\theta_{\rm max}\ll \theta_{\rm div}$, one finds

\begin{equation}
\frac{\Delta P_d}{P_{\rm inc}} \approx \frac{k}{\pi \omega_0^2}\left[\mathfrak{I}\left(\alpha\right)\frac{1}{1+z_p^2/z_R^2} + \mathfrak{R}\left(\alpha\right)\frac{z_p/z_R}{1+z_p^2/z_R^2}\right].\label{eq:PdRayleigh}
\end{equation}
Again, the GLMT expression simplified to the expected result for an induced dipole. To see this, consider the far-field of the incidence beam on the optical axis with which the dipole field interferes. Using the paraxial Gaussian field amplitude in Eq.\ \eqref{eq:IncField}, now in a particle-centered coordinate system, one has 
\begin{equation}
\mathbf{E^i}\left(z\right)\rightarrow -\mathbf{\hat{e}_x} E_0 i z_R \frac{\exp\left(-ik\left[z+z_p\right]\right)}{z},\label{eq:IncFarField}
\end{equation}
where $i$ accounts for the limiting Gouy phase $\phi_G\rightarrow \pi/2$. For the interference, in contrast to the scattering cross-section, now the phase $\phi_{\rm sca}$ of the scattered dipole field in Eq.\ \eqref{eq:ScaField} relative to the local incidence field becomes important. While it is tempting to assume a constant phase shift of the scattered field of $+\pi/2$ according to the Kirchhoff diffraction integral in combination with Babinet's principle \cite{HwangMoerner2007}, a careful analysis of subwavelength diffraction by circular apertures \cite{Lee2013} in the spirit of Bethe's original work shows that such an approach is invalid. Instead, this relative phase is determined by the complex value of the polarizability in Eq.\ \eqref{eq:alpha}. The interference which determines the relative transmission signal finally gives
\begin{equation}
\frac{\Delta P_d}{P_{\rm inc}} \approx \frac{2\,\mathfrak{R}(\mathbf{E^{i*}}\!\cdot\! \mathbf{E^s})}{|\mathbf{E^i}|^2} = \frac{-k}{\pi\omega_0^2}\mathfrak{R}\left(i \alpha \frac{1}{1-iz_p/z_R}\right)\label{eq:DipoleInterferenceForward},
\end{equation}
which identically agrees with Eq.\ \eqref{eq:PdRayleigh}. These forms for the signal in the forward-direction are similar to the one discussed in Ref.\ \cite{HwangMoerner2007}. However, the expressions of that reference have an additional axially dependent multiplicative factor which is absent here, and the contribution of each form is now clearly connected to the particle property $\alpha$.

If the polarizability is purely imaginary, a simple dip in transmission caused by the absorption will be detected and is maximal for a particle positioned in the focus. As noted before \cite{HwangMoerner2007,Pototschnig2011}, a destructive interference results from the phase difference of more than $\phi_G-\phi_{\rm sca}\ge \pi/2$ in the forward direction between the scattered (far) field, radiating with a relative phase lag of $-\pi/2$, and the incidence far field which attains the rest of its total Gouy phase of $\pi$. If the particle is placed in the focus, this phase difference amounts to $\pi$. As Eq.\ \eqref{eq:RAextapprox} shows, for any collection angle the dip-like feature is the mere result of a partial detection of the absorption by the particle. It is therefore fundamentally different from the situation encountered for a two-level system \cite{Pototschnig2011}. The fact that the transmission signal mirrors the axial dependence of the absorbed power $\propto \sigma_{\rm abs}^\pi$ is here found to originate from the interplay between 1) the reduction of the dipole-strength $\mathbf{p}\propto \mathbf{E^i}\left(z_p\right)$ and 2) the exact far field phase-difference realized at each position.

If the polarizability is purely real-valued, the phase-difference attained relative to the incidence beam in the far field depends on the particle position within the beam. The interference of the scattered and the incident field causes either a decrease or an increase of transmission detectable in the forward direction caused by an energy-flux redistribution. Only for no offset, i.e.\ if the particle is placed in the focus, the relative phase-shift in the far field is exactly $\phi_{G}-\phi_{\rm sca}=\phi_G= \pi/2$, leading to a cancellation of the time-averaged interference over one optical period $2\pi/\omega$, whereby the detected relative signal vanishes.



In general, a metallic nanoparticle will have both real and imaginary parts of its polarizability. Accordingly, a zero-crossing of the relative transmission signal is predicted at a finite particle displacement of $z_p^0 \approx -z_R \mathfrak{I}\left(\alpha\right)\! /\mathfrak{R}\left(\alpha\right)$. This allows the determination of a nano particle's absolute position relative to the beam waist if its material properties are known, i.e.\ to determine the exact amount of energy absorbed. The maximal signals are found at $z_p^{\pm}= z_p^0 \pm z_R |\alpha| /\mathfrak{R}\left(\alpha\right)$ with 
\begin{equation}
\frac{\Delta P_d\left(z_p^{\pm}\right)}{P_{\rm inc}} \approx \frac{k\left[\mathfrak{I}\left(\alpha\right) \pm |\alpha|\right]}{2\pi\omega_0^2}.\label{eq:DPd0} 
\end{equation}
This means, assuming equal magnitudes in polarizabilities, that a resonant particle will cause a dip at $z_p=0$ twice as large as compared to the maximum signal for an off-resonant particle at $z_p=\pm z_R$, see Fig.\ \ref{Fig:Azp}b). This is a consequence of the reduced intensity at the respective particle offsets. 

\begin{figure}[bt]
\centerline{\includegraphics [width=\columnwidth]{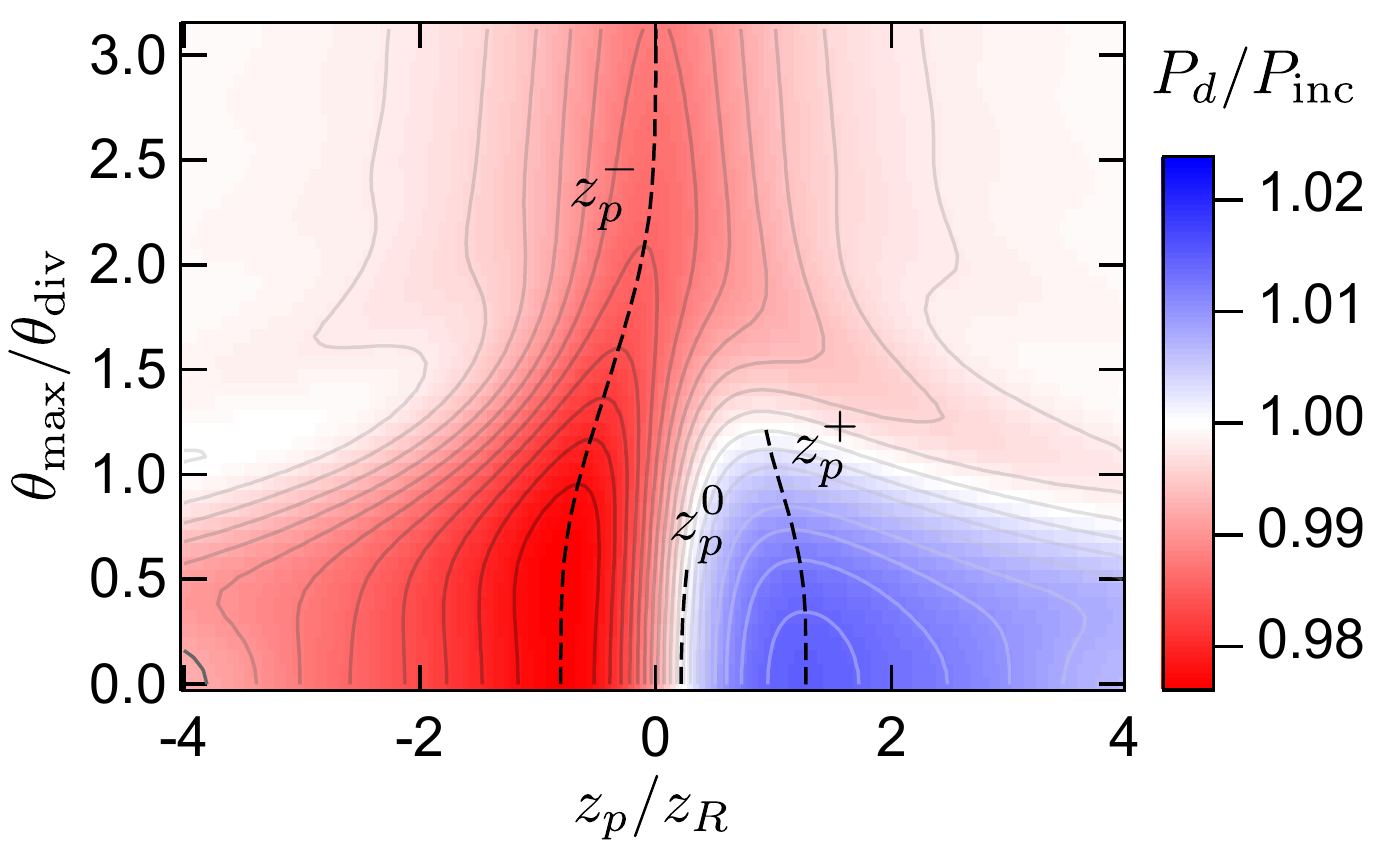}}
\caption{Axial scans of a $R=30\,\rm nm$ gold nanoparticle with a complex-valued polarizability $\alpha=\left[7.56 - 1.67i\right] \times 10^5\, {\rm nm}^3$ via $a_1$ for different collection angles $\theta_{\rm max}$, beam parameters as in Fig.\ \ref{Fig:Redistribution}. The signal is dispersive in the near-forward direction signifying the predominant real part of $\alpha$. The peak values are determined by eq.\ \eqref{eq:DPd0}. For large collection angles compared to the beam's angle of divergence the signal becomes dip-like due to the imaginary part of $\alpha$ and is determined by eq.\ \eqref{eq:pwNaive} (eq.\ \eqref{eq:scat} for a perfect dielectric).}\label{Fig:ScansThetamax} 
\end{figure}

\paragraph{Transmission at finite collection angles}
Arguably, the previous discussion could have been done using the concept of an induced dipole only. However, inconsistencies arise when such a treatment is extended to finite collection angles. For instance, a finite interference signal is then predicted for perfect dielectric particles and large collection angles \cite{Pralle1999}, which is inconsistent with the notion of an energy redistribution without absorption. The difficulties which arise if one attempts to take the paraxial Gaussian beam along with a dipole field are reminiscent of the intricacies encountered by M.\ Berg \cite{Berg2008} for the fractional extinction even in plane wave scattering. Now, the strength of the GLMT lies in its ability to evaluate the interference contribution for such finite collection angles.

In the intermediate regime of a collection angle which is of the order of the beam's angle of divergence $\theta_{\rm max}\sim \theta_{\rm div}$, the signal behaves in a peculiar way: No longer does the relative transmission signal follow the absorbed power even for a resonant particle. The axial signal deviates from a Lorenzian profile and broadens up, see Fig.\ \ref{Fig:Azp}a). Likewise, the disperse signal of a perfect dielectric becomes more complex and exhibits more than a single zero-crossing. A quick glance at Fig.\ \ref{Fig:ScansThetamax} is enough to see that the signal shape and magnitude change significantly over an angular domain which corresponds to twice the beam's angle of divergence. Even the sign changes from positive to negative for $z_p >0$, see also Fig.\ \ref{Fig:Redistribution}b). An accurate description of the signal then necessarily requires Eq.\ \eqref{eq:RAextapprox}. It is in this angular domain where typical transmission setups operate \cite{Muskens2008}.

Considering now the limit of a large numerical aperture detection, such that $\theta_{\rm max}\gtrsim 2\theta_{\rm div}$, the imaginary part of the complex axial shape function vanishes, $\mathfrak{I}\left(A\right)=0$, while the real part simply becomes $\mathfrak{R}\left(A\right)=1/\left[1+z_p^2/z_R^2\right]$, see Fig.\ \ref{Fig:Azp}a,c). Further, the incidence cross-section to be used for normalization is then close to its limiting value $\sigma_{\rm inc}^{\pi/2}=\pi \omega_0^2/2$, corresponding to $P_{\rm inc}$ being the total power of the incidence beam. Then,
\begin{equation}
\frac{\Delta P_d}{P_{\rm inc}} \approx \frac{1}{1+z_p^2/z_R^2}\frac{2k\,\mathfrak{I}\left(\alpha\right)}{\pi\omega_0^2} < 0. \label{eq:pwNaive}
\end{equation}
This value corresponds to a plane-wave approximation, $\sigma_{\rm ext} I_0/P_0$ with $I_0$ being the intensity at the particle's coordinate, which one could also have tentatively identified with the paraxial limit discussed before. However, this result for a resonant particle is too large by a factor of two as compared to the correct value in the forward direction given in eq.\ \eqref{eq:DPd0}. The reason for this was encountered in Fig.\ \ref{Fig:Redistribution}b) or Fig.\ \ref{Fig:Azp}b,c): While most of the power of the incidence beam is already contained within its angle of divergence, the interference representing an energy redistribution and accounting for the absorption occurs on an angular scale which is about twice that large. Therefore, the relative transmission for a resonant particle is smaller in the forward direction. As $\sigma_{\rm ext}\rightarrow \sigma_{\rm ext}^\pi \approx \sigma_{\rm abs}^\pi$ for small particles, it is the entire energy $P_{\rm abs}$ absorbed by the particle which is detected as missing in the beam propagation direction. For equal illumination and collection angles, $\theta_{\rm max}\sim \theta_{\rm div}$, as realised for two equal objectives used in a SMS setup, a systematic difference of the order of $25\%$ for extracted absorption cross-sections is expected.

However, for a perfect dielectric particle the extinction as approximated in eq.\ \eqref{eq:RAextapprox} vanishes. To correctly recover the extinction cross-section in this limit, higher-order terms in the expansion of the dipolar Mie coefficient $a_1\left(\alpha\right)$ must be considered. Alternatively, the energy balance may be used in the form of $-\sigma_{\rm abs}^\pi=\sigma_{\rm sca}^\pi-\sigma_{\rm ext}^\pi$. The fractional extinction cross-section saturates already for $\theta_{\rm max}\approx 2\theta_{\rm div}$ at the corresponding value $\sigma_{\rm ext} \rightarrow \sigma_{\rm sca}^\pi+\sigma_{\rm abs}^\pi$, see also Fig.\ \ref{Fig:Redistribution}b). While the absorption was considered before in eq.\ \eqref{eq:pwNaive} and may vanish, it is then the scattering contribution which still remains. Thus, the additional (negative) relative signal for weakly or non-absorbing particles, now further including direct scattering in eq.\ \eqref{eq:Pd} which is of the same magnitude, reads
\begin{equation}
\frac{\Delta P_d}{P_{\rm inc}} \approx \frac{\sigma_{\rm sca}\left(\theta_{\rm max}\right) - \sigma_{\rm sca}^{\pi}}{\pi \omega_0^2/2}\propto \frac{-1}{1+z_p^2/z_R^2} \frac{k^4 |\alpha|^2}{\omega_0^2}.\label{eq:scat}
\end{equation}
The value of $k^3\alpha$ determines the collection angle at which this contribution starts to dominate, see Fig.\ \ref{Fig:Azp}b). The particle's contrast is now exceedingly small and $\propto |\alpha|^2\propto R^6$ instead of $\propto R^3$ under small angle detection conditions. Although only the scattering cross-section appears in the above result, it correctly accounts for the interference within the beam's forward direction. Indeed, due to destructive interference a cancellation which accounts for a full scattering cross-section occurs within twice the beam's angle of divergence, independent of the dipolar field structure of the scattered wave. Only the part due to the fractional scattering cross-section occurs over the full angular domain. Thereby, effectively the fraction of the non-collected scattered power within $\left[\theta_{\rm max},\pi\right]$ is detected missing (in addition to the absorption, Eq.\ \eqref{eq:pwNaive}). Thus, again, the large angle situation corresponds to the plane-wave calculation if one were to correctly identified the extinction cross-section with the fractional scattering cross-section for the backward-excluded angular domain (plus the absorption cross-section). In general, the relative transmission for a perfect dielectric decreases with increasing collection angle. Therefore, while an absorbing particle is best found using the largest detection aperture possible, a perfect dielectric is best detected at an offset and using a small collection angle of about $\theta_{\rm max}\approx 0.9 \theta_{\rm div}$, providing the best signal-to-noise ratio.


\paragraph{Generalization of the optical theorem and absorption measurements}
The previous discussions may be understood as a consequence of a generalization of the optical theorem: For plane-wave the theorem states that the entire physical information regarding the net-energy balance, that is the absorption and the scattering cross-section, is contained in the forward field amplitude, $\sigma_{\rm ext}^\pi = 4\pi k^{-2}\mathfrak{R}\left(S_1\left(0^\circ\right)\right)$. A corrected version was shown to hold also for particles located in the beam-waist at $z_p=0$ of a focused on-axis beam \cite{GouesbetOpticalTheorem}. In general, however, it is the phase of the scattered field relative to the far-field incidence beam which matters for the interference, such that no expression can be expected which only includes the forward scattering amplitude. The Gouy phase anomaly necessitates this complication. Instead, Eq.\ \eqref{eq:DipoleInterferenceForward} explicitly includes the NP position in the beam. Further, In the case of focused beam scattering, 'forward' now refers to twice the beam's angle of divergence, i.e.\ the range of the beam's propagation directions. Then indeed a quantity $\propto \sigma_{\rm abs}^\pi+\sigma_{\rm sca}\left(\left[\theta_{\rm max},\pi\right]\right)$, which signifies a meaningful 'extinction', is detected missing. It is distinct from both the total extinction cross-section $\sigma_{\rm ext}^\pi$, which is devoid of any physical meaning, as well as its fractional counterpart $\sigma_{\rm ext}\left(\theta_{\rm max}\gtrsim 2\theta_{\rm div}\right)$ for $\theta>2\theta_{\rm div}$, see Fig.\ \ref{Fig:Redistribution}b). This discrepancy disappears only in the plane-wave limit $\theta_{\rm div}\rightarrow 0$.


\subsection{Transmission signal for large particles}
\begin{figure}[tb]
\centerline{\includegraphics [width=\columnwidth]{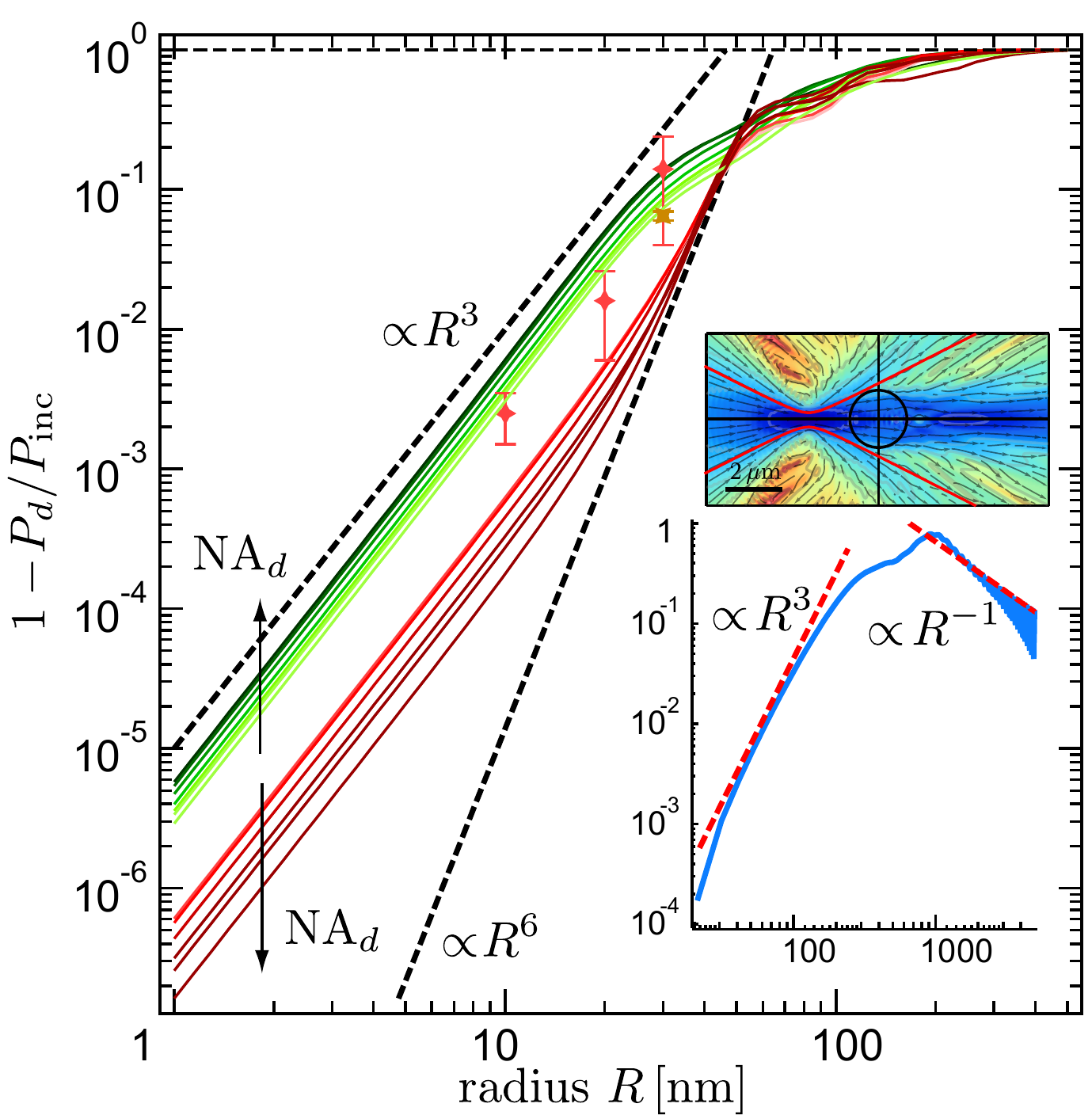}}
\caption{Maximum relative signals $-{\rm min}\left(\Delta P_d/P_{\rm inc}\right)$ of different sized AuNPs and Gaussian illumination. The red lines show the results for a beam waists $\omega_0 = 281\, \rm nm$ and $\lambda=635\,\rm nm$, $n_m=1.46$ for ${\rm NA}_d=\left\{0.1,0.3,0.5,0.75,1.0,1.3\right\}$ (from light to dark red). The red markers show experimental values for a single sample containing $R=10\, \rm nm$, $R=20\,\rm nm$ and $R=30\,\rm nm$ AuNPs for ${\rm NA}_d=0.75$. The green lines show the results for a beam waists $\omega_0 = 233\, \rm nm$ and $\lambda=532\,\rm nm$ for the same detection apertures (from light to dark green). The orange marker shows the experimental data for a beam waists of $\omega_0=315\,\rm nm$ for $\lambda=635\,\rm nm$. The inset shows the signal for transparent microspheres ($n_m=1.33$, $n_p=1.59$, $\omega_0=250\,\rm nm$, $\lambda=532\,\rm nm$, ${\rm NA}_d=0.05$), showing the limiting geometrical optics lensing behaviour $\propto R^{-1}$ at $z_p=-z_R$.}\label{Fig:RDep}
\end{figure}

According to eq.\ \eqref{eq:RAextapprox}, the depth of the relative transmission signals for small particles scales with their volume. Only for perfect dielectrics and large collection angles the dependence transitions to $\propto R^6$. Fig.\ \ref{Fig:RDep} shows this dependence for AuNPs in the size range of $R=1\,\rm nm$ up to particles of $R=0.5\,\mu \rm m$ on and off-resonance. The scaling with the volume of the particles holds well for particles in the Rayleigh size-regime but breaks down for larger particles as expected. The calculation of transmission signals via the analytical solutions to $\sigma_{\rm sca}$ and $\sigma_{\rm ext}$, i.e.\ eqs.\ \eqref{eq:AppendixSigmaExtOnAxis1} - \eqref{eq:AppendixSigmaScaOnAxis}, is only mildly more involved and correctly accounts for higher excited multipoles. For metallic particles whose size exceeds the beam waist of the incident beam, i.e.\ $R > \omega_0$, a saturation is reached indicating a complete extinction (no transmission). This is easily understood by a look at the near field under such conditions, see of Fig.\ \ref{Fig:interference}b). The entire beam is affected by the particle and the energy is removed from the forward direction entirely \cite{Locke1995}. The majority of the energy is retro-reflected via the scattered wave. A fraction determined by the Fresnel coefficient for reflection $r_n$ for normal incidence is also absorbed. For plane waves, this amounts to the geometrical limit $\sigma_{\rm ext}^\pi\rightarrow 2 \mathcal{A}$ and $\sigma_{\rm abs}^\pi\rightarrow \mathcal{A}\, \left[1-|r_n|^2\right]$ with the area being the particles cross-sectional area $\mathcal{A}=\pi R^2$. For focused illumination the area is determined by the beam spot size, $\mathcal{A}=\sigma_{\rm inc}^{\pi/2}= \pi \omega_0^2/2$. A transparent particle will show a limiting lens-like behaviour and accordingly $\Delta P_d/P_{\rm inc}=2z_p/f$ \cite{SelmkeABCD}, with a focal length $f=R\left[n_p/n_m\right]/(2\left[n_p/n_m-1\right])$.

\subsection{Off-axis scattering}
\begin{figure}[tbh]
\centerline{\includegraphics [width=\columnwidth]{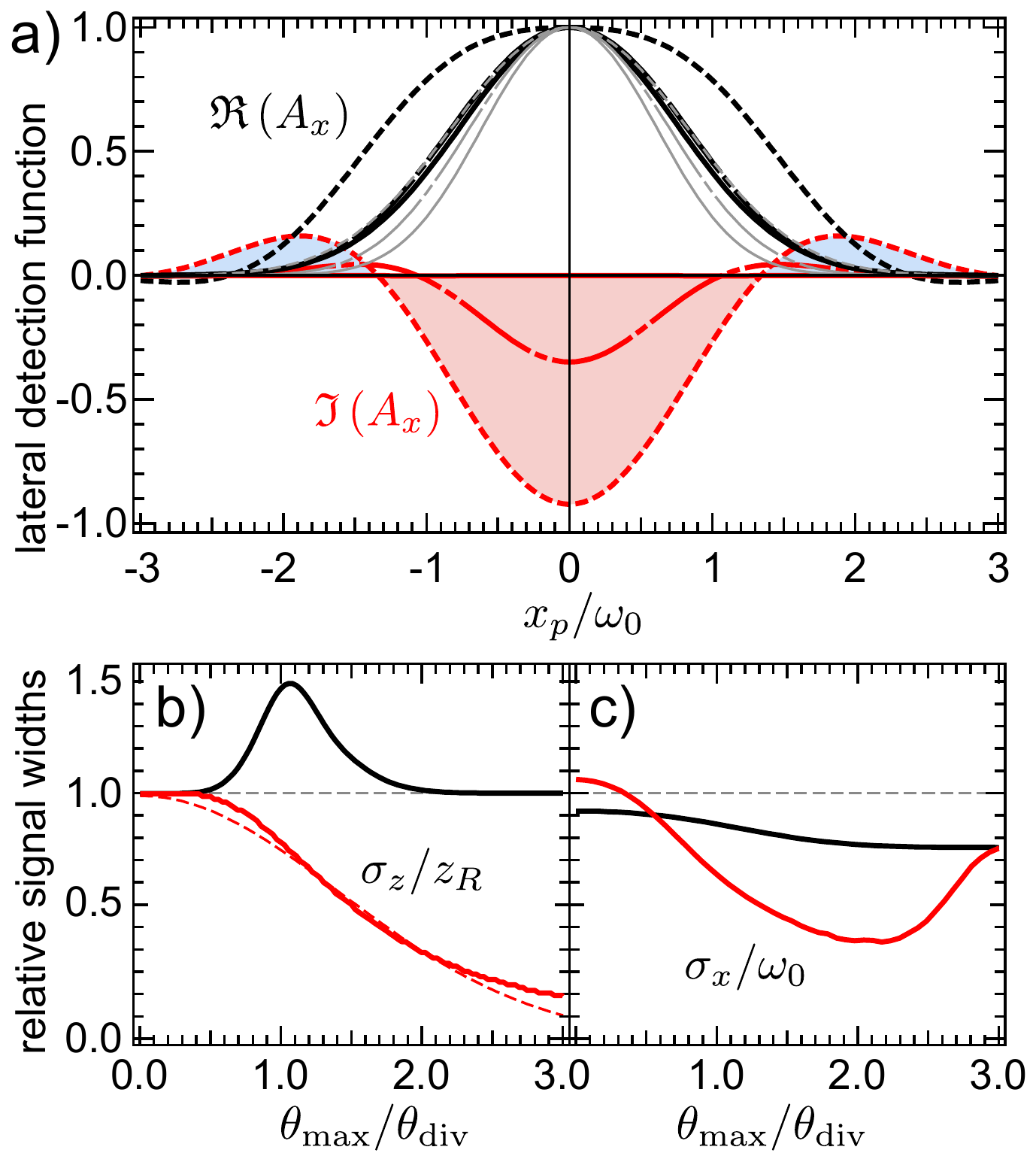}}
\caption{\textbf{a)} Real (black and gray) and imaginary part (red) of the lateral detection function $A_x\left(x_p\right)$. Dashed, solid-dashed and solid lines correspond to $\theta_{\rm max}=0,\theta_{\rm div}$ and $\pi/2$, respectively. The thick lines are for $z_p=-z_R$, the thin gray lines for $z_p=0$ (imaginary parts are zero here). \textbf{b)} Relative axial widths $\sigma_z/z_R$ and \textbf{c)} lateral widths $\sigma_x/\omega_0$ of a $R=30\,\rm nm$ resonant and perfect dielectric particles, beam parameters as in Fig.\ \ref{Fig:Redistribution}. The axial widths are defined as $\sigma_z=\left[z_p^{+}-z_p^{-}\right]/2$ for the dispersive case (red) and from a Lorenzian fit $\propto 1/\left[1+z_p^2/\sigma_z^2\right]$ for the dip-like case (black line). The lateral widths are obtained from a Gaussian fit with $\propto \exp\left(-x^2/\sigma_x^2\right)$ (at $z_p=z_p^{-}$ and $z_p=0$, respectively).}\label{Fig:ScanWidths}
\end{figure}

\begin{figure}[bth]
\centerline{\includegraphics [width=\columnwidth]{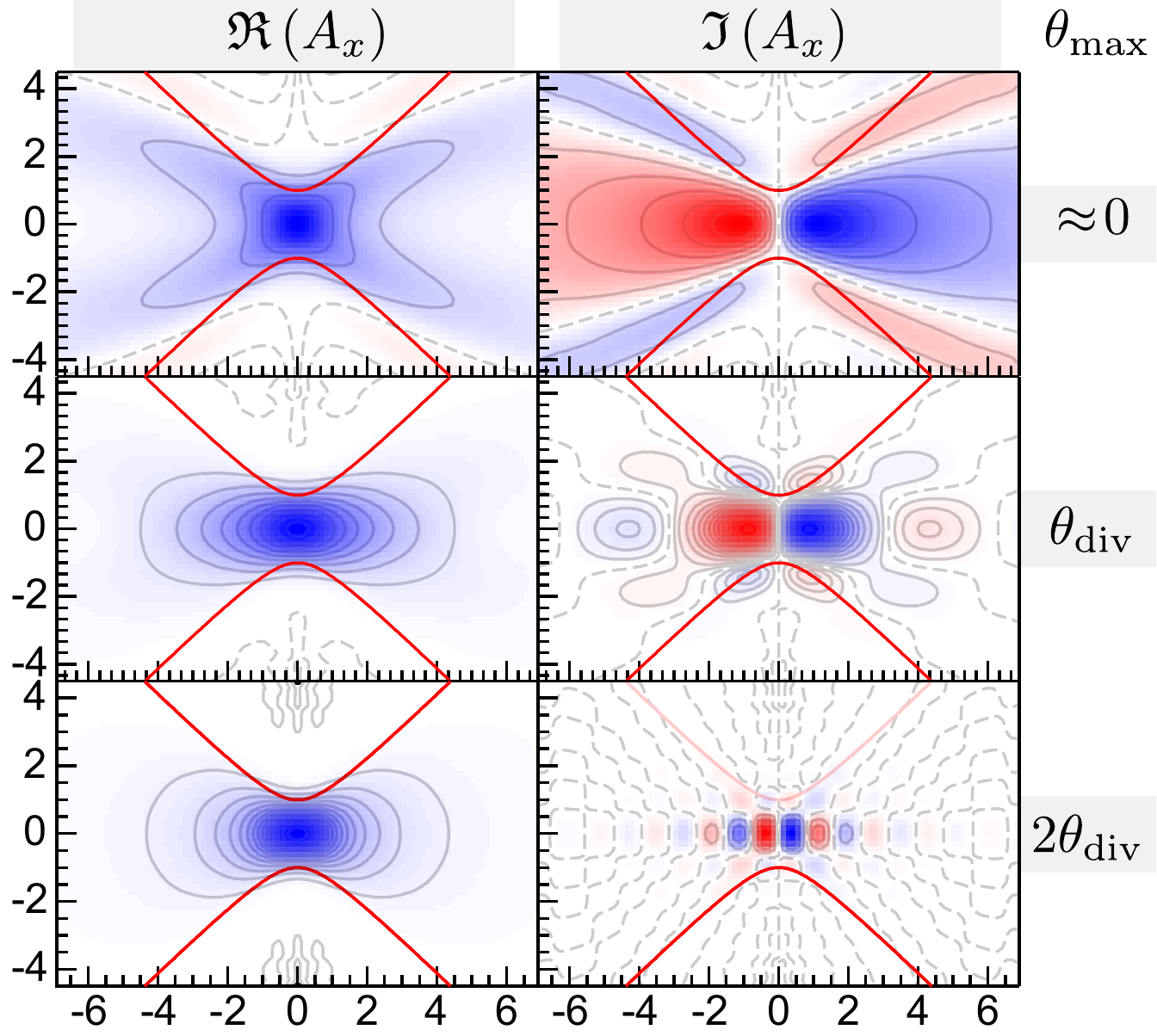}}
\caption{Real (left column) and imaginary part (right column) of the detection function $A_x\left(x_p/\omega_0,z_p/z_R\right)$ ($A_x\left(0,z_p\right)=A$). The colour-scale amplitudes are shown in Fig.\ \ref{Fig:Azp}b). The rows correspond to the collection angles $\theta_{\rm max}=0,\theta_{\rm div}$ and $2\theta_{\rm div}$, respectively (top to bottom). The bottom left plot mirrors the incident beam's intensity profile $\propto |\mathbf{E^i}|^2$.}\label{Fig:ScanZX}
\end{figure}

The previous discussions assumed that the beam waist $\omega_0$ is known, e.g.\ for extracting the absolute value of the complex-valued polarizability via eq.\ \eqref{eq:DPd0} and \eqref{eq:pwNaive}. In order to see how the beam waist $\omega_0$ can be measured in transmission microscopy we continue to consider such transmission signals in the GLMT. In the most general case of off-axis scattering the evaluation of the two integrals for the fractional cross-sections and thereby collected powers involves integrands which are the product of two double-sums, while the incidence cross-section eq.\ \eqref{eq:sigmainc} can still be used. The corresponding BSCs $g_{n,{\rm TM,TE}}^m$ for Gaussian beams can be found in Ref.\ \cite{DoicuGnm,Gouesbet2011} and many others. Other beams may be considered by using the corresponding BSCs. The fractional cross-sections for arbitrary illumination \cite{SelmkeACSNano,SelmkeABCD} may then again be reduced to a form in which only tabulated integrals appear, see appendix Eq.\ \eqref{eq:AppendixSigmaScaOffAxis} and \eqref{eq:AppendixSigmaExtOffAxis}. For small particles the following may be found
\begin{align}
\sigma_{\rm ext}&=-k\left[\mathfrak{I}\left(\alpha\right)\mathfrak{R}\left(A_x\right)+\mathfrak{R}\left(\alpha\right)\mathfrak{I}\left(A_x\right)\right].\label{eq:SigmaExtOffaxisRayleigh}
\end{align}
The complex-valued function $A_x\left(x_p\right)$ for any axial coordinate $z_p$ depends on the lateral particle offset $x_p$, the beam parameters as encoded in the BSCs and the collection angle $\theta_{\rm max}$. It determines the lateral shape of the relative transmission signal. While the real part shows a simple dip, the imaginary part shows again a dispersive lateral feature, see Fig.\ \ref{Fig:ScanWidths}a). The weight of each contribution is determined by the real and imaginary part of the polarisability, much in the same way as before for the axial shape, cf.\ Eq.\ \ref{eq:RAextapprox}. Only for $z_p=0$ are the lateral transmission scans are well fitted by a Gaussian function. For larger particle offsets and especially for small collection angles the shape deviates significantly from the beam's intensity distribution even for resonant particles, see the thick black dashed line in Fig.\ \ref{Fig:ScanWidths}a and the contours in Fig.\ \ref{Fig:ScanZX} (top left). The two constituting two-dimensional detection volumes are shown in Fig.\ \ref{Fig:ScanZX} and summarise the axial and lateral signal shape characteristics. For the gaussian beam considered here, particularly complex interference patterns are observed for large collection angles and perfect dielectrics. The above description remains valid also for the lateral coordinate $y_p$ perpendicular to the plane of polarisation. While the pattern largely remains the same, subtle differences appear for tight focusing (large $s$), such as a narrowing effect and more pronounced interference fringes \cite{SelmkeACSNano}. The reduced width of the transmission signal pattern along the $y$-direction reflects the slightly elliptic intensity distribution $|\mathbf{E^i}|^2\left(x,y\right)$ in real-space of the focused Davis beam \cite{LockTightFocusing}.

\subsection{The spatial extent of the signal}
As expected for light microscopy setups, the spatial extent of a discernible signal around the focal point increases for decreasing collection angle $\theta_{\rm max}$. Fig.\ \ref{Fig:ScanWidths}b) shows the evolution of the axial extend of the signal. The red curve depicts the case of a perfect dielectric particle, where the width is defined via the distance of the two peaks comprising the dispersive signal shape. It is seen to decrease monotonically, approximately following $\exp\left(-\theta_{\rm max}^2/4\theta_{\rm div}^2\right)$ (dashed line). The black curve shows the case of a resonant particle, where the width is defined via an effective Rayleigh range from a fit to the dip-like scan. A maximum in the axial width is seen for the case of $\theta_{\rm max}\approx \theta_{\rm div}$. The shape of the signal was already depicted in Fig.\ \ref{Fig:Azp}a) for the case of maximum and minimum width.

Evaluating the relative transmission signal at different lateral offsets $x_p$ allows the determination of the lateral extent of the signal. Fitting a Gaussian $\propto \exp\left(-x_p^2/\sigma_x^2\right)$ at the axial position $z_p^{-}$ of least transmission yields the curves shown in Fig.\ \ref{Fig:Azp}c). In both cases, for the perfect dielectric (red) and a resonant particle (black), the width is about $\sigma_x\approx \omega_0$, as expected for an interference signal. This corresponds to the lateral dependence of the electric field amplitude of the incident beam in the focal region, eq.\ \eqref{eq:IncField}. Therefore, measuring a lateral scan for a small collection angle allows for a robust determination of the beam-waist $\omega_0$ necessary for the extraction of absolute values for the polarizability $\alpha$. For a resonant particle the width decays with increasing collection angle until saturating at a lower value $\sigma_x \approx \omega_0/\sqrt{2}$, corresponding to the incidence beam intensity profile. The signal width for a perfect dielectric transitions through lower values until it reaches the same saturation value, indicating again that the signal is finally due to scattering only and follows the intensity profile.

%

\subsection{Spectral shift for finite particle offsets}
\begin{figure}[tb]
\centerline{\includegraphics [width=\columnwidth]{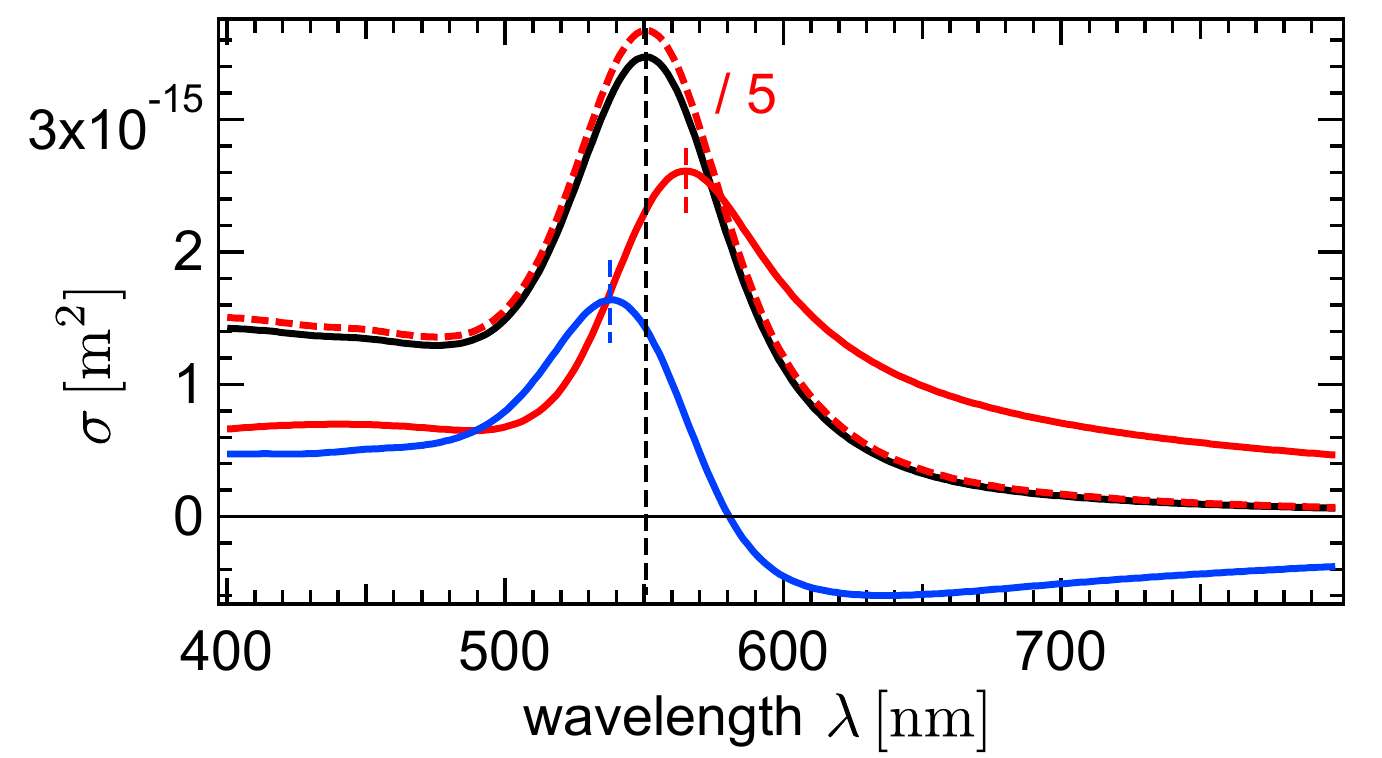}}
\caption{Plane-wave LM theory extinction spectrum for a $R=30\,\rm nm$ AuNP (red-dashed, scaled by $1/5$) and the measurable signal $\sigma_{\rm ext}-\sigma_{\rm sca}$ via eq.\ \eqref{eq:PdRel}, from the GLMT formalism. Beam parameters are $\omega_0=300\,{\rm nm}\times \lambda/635\,{\rm nm}$, $n_m=1.46$, ${\rm NA}_d=0.3$. The thin lines indicate the measurable signal for $z_p=0.57\,\mu\rm m$ (blue) and $z_p=-0.57\,\mu\rm m$ (red), showing apparent resonance shifts of $-12\,\rm nm$ and $+15\rm nm$, respectively. The Lorenzian peak widths are reduced by $\approx 18$\% and $16$\%, respectively.}\label{Fig:Spectrum}
\end{figure}

While the placement of a NP on the optical axis is easily done by adjusting for the maximum relative signal, it is clear that such a configuration in general does not correspond to a placement of the particle at the center of the beam waist $z_p=0$ for any real part of the polarizability $\mathfrak{R}\left(\alpha\right)\ne 0$.
Fig.\ \ref{Fig:Spectrum} shows the result of a spectrum as it would be extracted from transmission for a small AuNP, i.e.\ computed without further processing from the measured signal via $-\sigma_{\rm inc} \Delta P_d/P_{\rm inc}$. For the situation depicted, the ratio of the collection angle to the divergence angle of the incident beam has the constant value $\theta_{\rm max}/\theta_{\rm div}=0.45$. As a consequence, the extracted extinction cross-section is only a fraction of the limiting value for plane-wave scattering. The form of the curve and its resonance peak appears unchanged for a particle placed at $z_p=0$, since here the energy redistribution does not contribute to the signal. Accordingly, the signal follows the absorption which is (partially) detected, whereby the form of the extinction spectrum is correctly recovered for the small particle. For a placement below or above the focus an apparent resonance shifts of the order of $\pm 10\,\rm nm$ occur along with a reduction of the peak width. The observed effect for the gold NP is a result of the growing importance of the energy redistribution effect $\propto \mathfrak{R}\left(\alpha\right)$ with increasing wavelength. Even for $\theta_{\rm max}\sim \theta_{\rm div}$ shifts of a few nm are expected. These resonance shifts for finite offsets disappear only when the collection angle well exceeds the divergence angle. The energy redistribution also leads to an apparent negative extinction for $z_p>0$ at wavelengths beyond the resonance peak. These effects may be of importance if a white-light transmission setup is used in which chromatic aberrations play a role. An additional homogeneous broadening of a recorded resonance peak is then expected and may only be avoided by ensuring that the collection angle fulfils $\theta_{\rm max}> 2\theta_{\rm div}$.

\section{Conclusion}
Within the generalized Lorenz-Mie framework a convenient set of analytical expressions have been presented which accurately describe the signal in a transmission microscopy system. Measuring the relative transmission for the two limiting cases of small ($\theta_{\rm max}\ll \theta_{\rm div}$) and large detection angle ($\theta_{\rm max}\gtrsim 2\theta_{\rm div}$), i.e.\ determining eq.\ \eqref{eq:DPd0} and eq.\ \eqref{eq:pwNaive}, allows the determination of the real and imaginary part of a small particle's complex-valued polarizability. In reverse, these expressions predict the axial signal shapes for a given polarizability and collection angle, thereby generalizing the results of Ref.\ \cite{HwangMoerner2007} to the practical regime of finite apertures. Further, transmission microscopy was seen to provides a good estimate for the plane-wave extinction cross-section of small particles under conditions where the collection angle exceeds twice the beam's divergence angle. For perfect dielectrics, the contrast was seen to vanish due to the intrinsic nature of an energy redistribution over the same angular scale. Since our previous studies have shown that the corrected Gaussian beam well describes even tightly focused beams, the results are expected to show the general features of any tightly focused beam.  


\clearpage
\newpage
\onecolumngrid
\section{Appendix}

\subsection{Spherical field components of $\mathbf{E^{i}}$ and $\mathbf{H^{i}}$\label{appendix:ExplicitFields}}
The following expressions represent the electromagnetic field of the incident beam for on-axis scattering \cite{Gouesbet1988}:
%

\begin{align}
E_r^i &= -E_0 \cos\phi \sum_{n=1}^{\infty} f_n \left[\frac{1}{k}\left(r j_n\right)''+ k r\,j_n\right]\sin\left(\theta\right) \Pi_n\nonumber,\\
E_\theta^i &= \frac{E_0}{r} \cos\phi \sum_{n=1}^{\infty} f_n \left[-\frac{1}{k}\left(r j_n\right)'\tau_n+i r j_n \Pi_n\right]\nonumber,\\
E_\phi^i &= \frac{E_0}{r} \sin\phi \sum_{n=1}^{\infty} f_n \left[\frac{1}{k}\left(r j_n\right)'\Pi_n-i r j_n \tau_n\right],\nonumber
\end{align}
and the magnetic field strength
\begin{align}
H_r^i &= -H_0 \sin\phi \sum_{n=1}^{\infty} f_n \left[\frac{1}{k}\left(r j_n\right)''+ k r\,j_n\right]\sin\left(\theta\right) \Pi_n\nonumber,\\
H_\theta^i &= \frac{H_0}{r} \sin\phi \sum_{n=1}^{\infty} f_n \left[i r j_n\Pi_n-\frac{1}{k}\left(r j_n\right)' \tau_n\right]\nonumber,\\
H_\phi^i &= \frac{H_0}{r} \cos\phi \sum_{n=1}^{\infty} f_n \left[i r j_n\tau_n-\frac{1}{k}\left(r j_n\right)' \Pi_n\right].
\end{align}
Herein, the arguments of the spherical Bessel functions $j_n$ and $h^{(2)}_n$ are $kr$, and $f_n=i^{n+1}\left(-1\right)^{n} N_n\, g_n$. These expressions have been used in their asymptotic form for the fractional cross-section evaluations. To see that indeed the angular function $M\left(\theta\right)$ is related to the incidence field one may express the derivative of any spherical Bessel function via recurrence relations $j_{n}'=j_{n-1}-\left[n+1\right] j_n/z$ and $j_{n-1}+j_{n+1}=\left[2n+1\right] j_n /z$, as suggested by J.A.\ Lock \cite{Locke1995}. Successive application of these shows that $(z j_n\left(z\right))' = z\left[j_{n-1}-n j_n/z\right]$ and $(z j_n)''=n\left(n+1\right)j_n/z - z j_{n}$. The incidence field in the near-forward direction may then be simplified using $\Pi_n \approx \tau_n$ and the relations $\mathbf{\hat{e}_x}=\cos\left(\theta\right)\mathbf{\hat{e}_\theta}-\sin\left(\theta\right)\mathbf{\hat{e}_\phi}$ as well as the asymptotic forms $j_n\rightarrow \sin\left(kr-n\pi/2\right)/kr$ and consequently $j_n+i j_{n-1}\rightarrow i^n i \exp\left(-ikr\right)/kr$. One finally arrives at:

\begin{equation}
\mathbf{E^i}\approx -\mathbf{\hat{e}_x} \frac{iE_0}{kr}\exp\left(-ikr\right)\frac{1}{2} \sum_{n=1}^{\infty}N_n g_n \left[\Pi_n+\tau_n\right]\label{Efieldj},
\end{equation}
for $\theta\ll 1$ and $kr\gg 1$. Indeed, the function $M\left(\theta\right)$, eq.\ \eqref{eqnM}, introduced in the far field extinction flux is closely related to the incident beam, i.e.\ one may write $\mathbf{E^i}\left(\mathbf{r}\right)\approx -\mathbf{\hat{e}_x} \left[iE_0M\left(\theta\right)\!/2\right]\exp\left(-ikr\right)/kr$. For the Gaussian BSCs the sum appearing in $M$ may be approximated by the corresponding integral $\int_{1}^{\infty}\mathrm{d}n$ to find $M\left(\theta\right)\approx 4\theta_{\rm div}^{-2}Q^{-1}$, such that in the far field 
\begin{equation}
\mathbf{E^i}\left(z\right)\rightarrow -\mathbf{\hat{e}_x} E_0 z_R \frac{\exp\left(-ik\left[z+z_p\right]+i\pi/2\right)}{z}.\label{Appendix:IncFarField}
\end{equation}
This expression describes the beam relative to a particle-centered coordinate system, and is consistent with the paraxial Gaussian field amplitude eq.\ \eqref{eq:IncField} relative to its beam-waist upon writing $1/\left[1-i z/z_R\right]\rightarrow \exp\left(i\pi/2\right) z_R/z$. The Davis beam thus agrees with the paraxial Gaussian beam approximation in the far field and on the optical axis, whereby the simple dipole concept works in this configuration. For $\theta_{\rm max}\ne 0$ the paraxial beam fails to account for components in the non-perpendicular directions.

\subsection{Spherical field components of $\mathbf{E^{s}}$ and $\mathbf{H^{s}}$\label{appendix:ExplicitFields}}
Similarly, the asymptotic form of the scattered field in the forward direction reads
\begin{equation}
\mathbf{E^s}\approx \mathbf{\hat{e}_x} \frac{iE_0}{kr}\exp\left(-ikr\right)\frac{1}{2}\sum_{n=1}^{\infty}N_n g_n \left[a_n+b_n\right]\left[\Pi_n + \tau_n\right] \label{Efieldsca},
\end{equation}
as can be inferred from the scattered field's spherical components \cite{Gouesbet1988} using the asymptotic form $h^{(2)}_n\rightarrow i^{n+1} \exp\left(-ikr\right)/kr$:
\begin{align}
E_r^s &= E_0 \cos\phi \sum_{n=1}^{\infty} f_n a_n\left[\frac{1}{k}\left(r h_n^{(2)}\right)''\!+ k r\,h_n^{(2)}\right]\sin\left(\theta\right) \Pi_n\nonumber,\\
E_\theta^s &= -\frac{E_0}{r} \cos\phi \sum_{n=1}^{\infty} f_n \left[\frac{-a_n}{k}\left(r h_n^{(2)}\right)'\!\tau_n+i b_n r h_n^{(2)} \Pi_n\right]\nonumber,\\
E_\phi^s &= -\frac{E_0}{r} \sin\phi \sum_{n=1}^{\infty} f_n \left[\frac{a_n}{k}\left(r h_n^{(2)}\right)'\Pi_n-i b_n r h_n^{(2)} \tau_n\right]\nonumber,
\end{align}
and the magnetic field strength
\begin{align}
H_r^s &= H_0 \sin\phi \sum_{n=1}^{\infty} f_n b_n\left[\frac{1}{k}\left(r h_n^{(2)}\right)''\!+ k r\,h_n^{(2)}\right]\sin\left(\theta\right) \Pi_n\nonumber,\\
H_\theta^s &= -\frac{H_0}{r} \sin\phi \sum_{n=1}^{\infty} f_n \left[i a_n r h_n^{(2)}\Pi_n- \frac{b_n}{k}\left(r h_n^{(2)}\right)' \tau_n\right]\nonumber,\\
H_\phi^s &= -\frac{H_0}{r} \cos\phi \sum_{n=1}^{\infty} f_n \left[i a_n r h_n^{(2)}\tau_n- \frac{b_n}{k}\left(r h_n^{(2)}\right)' \Pi_n\right].
\end{align}
The resulting far field expressions when $r\rightarrow \infty$ have been used to compute the detectable power $P_d$. Twice the resulting total field's time-averaged Poynting vector has the radial, polar and azimuthal components (only real parts considered) $E_\theta^t H_\phi^{t*}-E_\phi^t H_\theta^{t*}$, $E_\phi^t H_r^{t*}-E_r^t H_\phi^{t*}$ and $E_r^t H_\theta^{t*}-E_\theta^t H_r^{t*}$, respectively. 

\subsection{Cartesian components of the Poynting vector}
The cartesian representation of the Poynting vector may be obtained from the spherical components as $S_z=\cos\left(\theta\right)S_r-\sin\left(\theta\right) S_\theta$ and $S_{x,y}=\sin\left(\theta\right) S_r +\cos\left(\theta\right) S_\theta$ in the $xz$-plane ($\phi=0$) or the $yz$-plane ($\phi=\pi/2$). These expressions have been used to generate the flow-field and intensity plots in Fig.\ 2.

\subsection{Fractional cross-sections (analytical)}
The following indefinite integrals for the quadratic products of Mie functions are listed in Ref.\ \cite{BabenkoBook} (B.40 - B.46, B.42 corrected, B.38):
\begin{align}
\left(I_{1}\right)_{n,n'}^m=&\int \left[m^2\Pi_n^m\Pi_{n'}^m+\tau_n^m\tau_{n'}^m\right]\sin\left(\theta\right)\mathrm{d}\theta=\left\{ \begin{array}{ll} 
\displaystyle \frac{-\sin^2\left(\theta\right)\left[n'\left(n'+1\right)\tau_n^m\Pi_{n'}^m - n\left(n+1\right)\Pi_n^m\tau_{n'}^m\right] }{n\left(n+1\right)-n'\left(n'+1\right)} & {\rm for}\,\,n\ne n'\\
\displaystyle -\frac{n\left(n+1\right)}{2n+1}\overline{H}_n^m + \sin^2\left(\theta\right)\Pi_n^m\tau_n^m & {\rm for}\,\, n=n'
\end{array}\right.\label{eq:I1}\\
\left(I_2\right)_{n,n'}^m=&\int \left[\Pi_n^m\tau_{n'}^m+\tau_n\Pi_{n'}^m\right]\sin\left(\theta\right)\mathrm{d}\theta=\sin^2\left(\theta\right)\Pi_n^m\Pi_{n'}^m.\label{eq:I2}
\end{align}
For on-axis scattering the angular functions are $\Pi_n^{m=1}\rightarrow\Pi_n$ and $\tau_n^{m=1}\rightarrow\tau_n$, and we write $\tau_n\left(\theta\right)$ for $\tau_n\left(\cos\left(\theta\right)\right)$. The recurrence relations for the contribution $H_n^m=\overline{H}_n^m\left(\theta_{\rm max}\right)-\overline{H}_n^m\left(0\right)$ to $\left(I_{1}\right)_{n,n}^1$ of the definite integrals corresponding to the above Eq.\ \eqref{eq:I1} in the range of $\left[0,\theta_{\rm max}\right]$ read:
\begin{align}
H_n \equiv &H_n^1,\quad H_1=\cos\left(\theta_{\rm max}\right)\left[3-\cos^2\left(\theta_{\rm max}\right)\right] - 2,\nonumber\\
H_{n+1} =& \frac{\left(n+2\right)H_n}{n}+\frac{\sin^2\left(\theta_{\rm max}\right)}{n}\left\{\cos\left(\theta_{\rm max}\right)\left[\left(n+2\right)\Pi_n^2+n \Pi_{n+1}^2\right] - 2\left(n+1\right)\Pi_n\Pi_{n+1}\right\}.
\end{align}
For off-axis scattering the corresponding recurrence relation generating all $H_n^{0\leq m\leq n}$ reads (B.43 - B.46):
\begin{align}
H_m^{m>1} = &\left[\left(2m-1\right)!!\right]^2\left[\cos\left(\theta_{\rm max}\right)\sin^{2m}\left(\theta_{\rm max}\right)+\frac{2m}{\left(2m-1\right)\left[\left(2n-3\right)!!\right]^2} H_{m-1}^{m-1}\right], \quad H_0^0 = \cos\left(\theta_{\rm max}\right) - 1,\\
H_{n+1}^{m\geq 0} =& \frac{\left(n+m+1\right)}{\left(n-m+1\right)}H_{n\geq m}^m+\frac{\sin^2\left(\theta_{\rm max}\right)\left\{\cos\left(\theta_{\rm max}\right)\left[\left(n+m+1\right)\left(\Pi_n^m\right)^2+\left(n-m+1\right) \left(\Pi_{n+1}^m\right)^2\right] - 2\left(n+1\right)\Pi_n^m\Pi_{n+1}^m\right\}}{\left(n-m+1\right)}.\nonumber
\end{align}
One finds for the fractional extinction cross-section $\sigma_{\rm ext}=\mathfrak{R}\left(\sigma_{\rm ext}^c\right)$ on-axis:
\begin{align}
\sigma_{\rm ext}^c=&\frac{\pi}{k^2}\sum_{n,n'=1}^{\infty}\,N_n N_{n'} \,g_n^{*} g_{n'} \,\left[a_{n'}+b_{n'}\right]  \int_{0}^{\theta_{\rm max}} \left[\left(\Pi_n\tau_{n'}+\tau_n\Pi_{n'}\right) + \left(\Pi_n\Pi_{n'} + \tau_n\tau_{n'}\right)\right]  \sin\left(\theta\right) \mathrm{d}\theta \label{eq:AppendixSigmaExtOnAxis0}\\
=&\frac{\pi}{k^2}\left\{-\sum_{n=2}^{\infty}\sum_{n'=1}^{n-1} \frac{N_n N_{n'} \left[g_n^{*}g_{n'} \left(a_{n'}+b_{n'}\right)+g_{n'}^{*}g_n \left(a_n+b_n\right)\right]}{n\left(n+1\right)-{n'}\left({n'}+1\right)}\sin^2\left(\theta_{\rm max}\right) \left[n'\left(n'+1\right)\tau_n\Pi_{n'}-n\left(n+1\right)\Pi_n\tau_{n'}\right] \right.\nonumber\\
&\left.+ \sum_{n=1}^{\infty} N_n^2 |g_n|^2 \left(a_n+b_n\right)\left[\sin^2\left(\theta_{\rm max}\right)\Pi_n\tau_n - \frac{n\left(n+1\right)}{2n+1}H_n\left(\theta_{\rm max}\right)\right] + R_{M}^{*}\left[R_a+R_b\right]\right\}.\label{eq:AppendixSigmaExtOnAxis1}\\ 
& {\rm with}\quad R_M=\sum_{n=1}^{\infty}N_n g_n\Pi_n\sin\left(\theta_{\rm max}\right),\,\, R_a=\sum_{n=1}^{\infty}N_n g_n\Pi_n a_n \sin\left(\theta_{\rm max}\right), \,\, R_b=\sum_{n=1}^{\infty}N_n g_n\Pi_n b_n \sin\left(\theta_{\rm max}\right).\label{eq:AppendixSigmaExtOnAxis2}
\end{align}
%
The fractional scattering cross-section on-axis reads:
\begin{align}
\sigma_{\rm sca}=&\frac{\pi}{k^2}\int_{0}^{\theta_{\rm max}} \!\!\!\sum_{n,n'=1}^{\infty}N_n N_{n'} g_n^{*}g_{n'} \left[\left(a_n^{*}a_{n'}+b_n^{*}b_{n'}\right) \left(\tau_n\tau_{n'} + \Pi_n\Pi_{n'}\right) + \left(a_n^{*}b_{n'} + b_n^{*}a_{n'}\right) \left(\tau_n\Pi_{n'} + \Pi_n \tau_{n'}\right)\right]  \sin\left(\theta\right) \mathrm{d}\theta\nonumber\\
=&\frac{\pi}{k^2}\left\{ -2\sum_{n=2}^{\infty}\sum_{n'=1}^{n-1} \frac{N_n N_{n'} \mathfrak{R}\left(g_n^{*}g_{n'} \left[a_n^{*}a_{n'}+b_n^{*}b_{n'}\right]\right)}{n\left(n+1\right)-n'\left(n'+1\right)}\sin^2\left(\theta_{\rm max}\right) \left[n'\left(n'+1\right)\tau_n\Pi_{n'}-n\left(n+1\right)\Pi_n\tau_{n'}\right] \right.\nonumber\\
&\left. + \sum_{n=1}^{\infty} N_n^2 |g_n|^2 \left(|a_n|^2+|b_n|^2\right)\left[\sin^2\left(\theta_{\rm max}\right)\Pi_n\tau_n - \frac{n\left(n+1\right)}{2n+1}H_n\left(\theta_{\rm max}\right)\right] + 2\mathfrak{R}\left(R_a^{*}R_b\right)\right\}\label{eq:AppendixSigmaScaOnAxis}
\end{align}
%

The exact beam's incidence cross-section $\sigma_{\rm inc}=\mathfrak{R}\left(\sigma_{\rm inc}^c\right)$ may be written as:
\begin{align}
\sigma_{\rm inc}^c
=&\frac{\pi}{2k^2}\int_{0}^{\theta_{\rm max}}\!\!\! \sum_{n,n'=1}^{\infty} N_n N_{n'} g_n g_{n'}^{*}\left[\left(\Pi_n \Pi_{n'} + \tau_n\tau_{n'}\right)\left(1-\left(-1\right)^{n+n'}\right) + \left(\Pi_n\tau_{n'} + \tau_n \Pi_{n'}\right)\left(1 + \left(-1\right)^{n+n'}\right)\right]\sin\left(\theta\right)\mathrm{d}\theta\nonumber\\
=&\frac{\pi}{k^2}\sin^2\left(\theta_{\rm max}\right) \left[\sum_{n=1}^{\infty}N_n^2 |g_n|^2 \Pi_n^2 + \sum_{n=2}^{\infty}\sum_{n'=1}^{n-1} N_n N_{n'} g_n g_{n'}^{*} \times \right.\nonumber\\
&\left.\qquad\qquad\qquad\qquad \left(\Pi_n\Pi_{n'}\left[1 + \left(-1\right)^{n+n'}\right]-\frac{n'\left(n'+1\right)\tau_n\Pi_{n'} - n\left(n+1\right)\Pi_n\tau_{n'}}{n\left(n+1\right)-n'\left(n'+1\right)}\left[1-\left(-1\right)^{n+n'}\right]\right)\right]\label{eq:AppendixInc}
\end{align}
For the total incidence cross-section the expression agrees with $\varkappa/k^2$, from Ref.\ \cite{Stout2012} upon using the special values for $\Pi_n\left(\pi/2\right)=-\cos\left(\pi \left[n + 1\right]/2\right) n!!/(n - 1)!!$ and $\tau_n\left(\pi/2\right)=-\left(n+1\right)\Pi_{n-1}$. For off-axis illumination the calculation is similar. The fractional cross-sections \cite{SelmkeACSNano,SelmkeABCD} now include the double-indexed BSCs for arbitrary illumination:
\begin{align}
\sigma_{\rm sca}&=\sum_{i=1}^{2}\frac{2\pi}{k^2}\int_{0}^{\theta_{\rm max}} \sum_{n,m} \left(X_i\right)_{n}^m \!\! \sum_{n'=n'_{|m|}}^{\infty}\!\!\!\! \left(X_i^{*}\right)_{n'}^m \sin\left(\theta\right)\, \mathrm{d}\theta\nonumber,\\
\sigma_{\rm ext}^c&=\sum_{i=1}^{2}\frac{2\pi}{k^2}\int_{0}^{\theta_{\rm max}}  \sum_{n,m} \left(Y_i\right)_{n}^m \!\! \sum_{n'=n'_{|m|}}^{\infty}\!\!\!\! \left(X_i^{*}\right)_{n'}^m \sin\left(\theta\right)\, \mathrm{d}\theta\label{eqnOffIS2},
\end{align}
where $\sigma_{\rm ext}=\mathfrak{R}\left(\sigma_{\rm ext}^c\right)$, the azimuthal order $m$ runs from $-n$ to $n$ and in the last sum $n'_{|m|}=\textnormal{max}(1,|m|)$. The following notation was hereby introduced:
\begin{align}
\left(Y_i\right)_n^m &=M_{i1} \tau_n^{|m|} + M_{i2}\Pi_n^{|m|}\\
\left(X_1\right)_n^m &=a_n M_{12} \Pi_n^{|m|} + b_n M_{11}\tau_n^{|m|}\\
\left(X_2\right)_n^m &=a_n M_{21} \tau_n^{|m|} + b_n M_{22}\Pi_n^{|m|}
\end{align}
with the recursively computable \cite{SelmkeACSNano} angular functions $\tau_n^{|m|}$ and $\Pi_n^{|m|}$ and the matrix elements
\begin{align}
\left(\begin{array}{cc} M_{11} & M_{12} \\
M_{21} & M_{22}
\end{array}\right) &=\,\, N_n \left(\begin{array}{cc} i g_{n,{\rm TE}}^m & m g_{n,{\rm TM}}^m  \\
g_{n,{\rm TM}}^m & i m g_{n,{\rm TE}}^m
\end{array}\right)
\end{align}
The scattering cross-section reads:
%
\begin{align}
\sigma_{\rm sca}=&\frac{2\pi}{k^2}\int_{0}^{\theta_{\rm max}} \sum_{n=1}^{\infty}\sum_{m=-n}^{n}\left(a_n M_{12} \Pi_n^{|m|} + b_n M_{11}\tau_n^{|m|}\right)\sum_{n'=n_{|m|}}^{\infty}\left(a_{n'}^{*} M_{12}^{'*} \Pi_{n'}^{|m|} + b_{n'}^{*} M_{11}^{'*}\tau_{n'}^{|m|}\right)\sin\left(\theta\right) \mathrm{d}\theta\nonumber\\
&+ \frac{2\pi}{k^2}\int_{0}^{\theta_{\rm max}} \sum_{n,m}\left(a_n M_{21} \tau_n^{|m|} + b_n M_{22}\Pi_n^{|m|}\right)\sum_{n'=n_{|m|}}^{\infty}\left(a_{n'}^{*} M_{21}^{'*} \tau_{n'}^{|m|} + b_{n'}^{*} M_{22}^{'*}\Pi_{n'}^{|m|}\right)\sin\left(\theta\right) \mathrm{d}\theta\nonumber\\
=&\frac{2\pi}{k^2}\int_{0}^{\theta_{\rm max}}\! \sum_{n,m,n'}\! N_n N_{n'} \left(\left[a_n a_{n'}^{*}g_{n,{\rm TM}}^{m}g_{n',{\rm TM}}^{m*} + b_n b_{n'}^{*} g_{n,{\rm TE}}^{m}g_{n',{\rm TE}}^{m*}\right] \left[m^2\Pi_n^{|m|} \Pi_{n'}^{|m|}+\tau_n^{|m|}\tau_{n'}^{|m|}\right]\right. \nonumber\\
 & \left. + im\left[-a_n b_{n'}^{*} g_{n,{\rm TM}}^m g_{n',{\rm TE}}^{m*}+ b_n a_{n'}^{*} g_{n,{\rm TE}}^{m}g_{n',{\rm TM}}^{m*}\right] \left[\Pi_n^{|m|}\tau_{n'}^{|m|}+\tau_n^{|m|}\Pi_{n'}^{|m|}\right] \right)\sin\left(\theta\right) \mathrm{d}\theta,\label{eq:AppendixSigmaScaOffAxis}
\end{align}
apart from numerical errors, $\sigma_{\rm sca}$ is purely real-valued. The extinction cross-section $\sigma_{\rm ext}=\mathfrak{R}\left(\sigma_{\rm ext}^c\right)$ reads:
\begin{align}
\sigma_{\rm ext}^c=&\frac{2\pi}{k^2}\int_{0}^{\theta_{\rm max}} \sum_{n=1}^{\infty}\sum_{m=-n}^{n}\left(M_{11} \tau_n^{|m|} + M_{12}\Pi_n^{|m|}\right)\sum_{n'=n_{|m|}}^{\infty}\left(a_{n'}^{*} M_{12}^{'*} \Pi_{n'}^{|m|} + b_{n'}^{*} M_{11}^{'*}\tau_{n'}^{|m|}\right)\sin\left(\theta\right) \mathrm{d}\theta\nonumber\\
&+ \frac{2\pi}{k^2}\int_{0}^{\theta_{\rm max}} \sum_{n,m}\left(M_{21} \tau_n^{|m|} + M_{22}\Pi_n^{|m|}\right)\sum_{n'=n_{|m|}}^{\infty}\left(a_{n'}^{*} M_{21}^{'*} \tau_{n'}^{|m|} + b_{n'}^{*} M_{22}^{'*}\Pi_{n'}^{|m|}\right)\sin\left(\theta\right) \mathrm{d}\theta\nonumber\\
=&\frac{2\pi}{k^2}\int_{0}^{\theta_{\rm max}}\! \sum_{n,m,n'}\!\!\! N_n N_{n'}\left(\left[a_{n'}^{*}g_{n,{\rm TM}}^{m}g_{n',{\rm TM}}^{m*} + b_{n'}^{*} g_{n,{\rm TE}}^{m}g_{n',{\rm TE}}^{m*}\right] \left[m^2\Pi_n^{|m|} \Pi_{n'}^{|m|}+\tau_n^{|m|}\tau_{n'}^{|m|}\right]\right. \nonumber\\
 & \left. + im\left[-b_{n'}^{*} g_{n,{\rm TM}}^m g_{n',{\rm TE}}^{m*}+ a_{n'}^{*} g_{n,{\rm TE}}^{m}g_{n',{\rm TM}}^{m*}\right] \left[\Pi_n^{|m|}\tau_{n'}^{|m|}+\tau_n^{|m|}\Pi_{n'}^{|m|}\right] \right)\sin\left(\theta\right) \mathrm{d}\theta.\label{eq:AppendixSigmaExtOffAxis}
\end{align}
In both cases the appearing integrals are of the form $\left(I_{1}\right)_{n,n'}^m$ and $\left(I_{2}\right)_{n,n'}^m$. For an efficient computation of a transmission scan $\Delta P_d\left(\mathbf{r_p}\right)/P_{\rm inc}$ these integrals should be evaluated only once and tabulated along with the scattering coefficients $a_n$ and $b_n$ and the normalization factors $N_n$. Only the $2\times \left(2n_{\rm max}+n_{\rm max}^2\right)$ BSCs $g_{n,{\rm TE}}^{m}\left(\mathbf{r_p}\right)$ and $g_{n,{\rm TM}}^{m}\left(\mathbf{r_p}\right)$ need to be evaluated at each position and the triple-sum over $n,m,n'$ with $n_{\rm max}+3n_{\rm max}^2+n_{\rm max}^3$ summands to be evaluated.

\subsection{Small particle limit (on-axis)}
For small particles  one finds that the fractional scattering cross-section becomes negligible, $\sigma_{\rm sca}\approx 0$. Starting from Eq.\ \eqref{eq:AppendixSigmaExtOnAxis0}, and effectively truncating the series in the scattering functions $S_{1,2}$ at their first term by setting $n'=1$, one finds for the fractional extinction cross-section (using $\tau_1=\cos\left(\theta_{\rm max}\right)$, $\Pi_1=1$):
%
\begin{align}
\sigma_{\rm ext} \approx&\, \frac{\pi}{k^2}\mathfrak{R}\left(a_1\left\{\frac{3}{4}\,|g_1|^2 \sin^2\left(\frac{\theta_{\rm max}}{2}\right)\left[15+8\cos\left(\theta_{\rm max}\right)+\cos\left(2\theta_{\rm max}\right)\right] \right. \right.\nonumber\\
  & + \left.\left. \frac{3}{2} g_1 \sin^2\left(\theta_{\rm max}\right) \sum_{n=2}^{\infty}\,N_n \,g_n^{*} \frac{n\left(n+1\right)\Pi_n\left[\cos\left(\theta_{\rm max}\right)+1\right]-2\left[\tau_n+\Pi_n\right]}{n\left(n+1\right)-2}\right\}\right)\label{eq:AppendixExtR}\\
  \equiv & \frac{\pi}{k^2}\mathfrak{R}\left(6 a_1 A\right) =\frac{\pi}{k^2}\left[\mathfrak{R}\left(a_1\right)\mathfrak{R}\left(6A\right) - \mathfrak{I}\left(a_1\right)\mathfrak{I}\left(6A\right)\right]=- k\left[\mathfrak{I}\left(\alpha\right)\mathfrak{R}\left(A\right) + \mathfrak{R}\left(\alpha\right)\mathfrak{I}\left(A\right)\right]\label{eq:AppendixExtR2}
\end{align}
wherein $A=A\left(z_p,\theta_{\rm div},\theta_{\rm max}\right)=\left\{\dots\right\}/6$ is a function of the particle coordinate $z_p$ within the beam via the coordinate-dependent beam shape coefficients $g_n\left(z_p\right)$, eq.\ \eqref{eqn:gn}. For small collection angles $\theta_{\rm max}\approx 0$ the sum in the above expression reduces to $\sum g_n^{*}\left(2n+1\right)$ since $\Pi_n\left(0\right)=\tau_n\left(0\right)=n\left(n+1\right)/2$. The term outside the sum $\approx |g_1|^2\theta_{\rm max}^2 9/2$ can then be reabsorbed into it. Using the Gaussian BSCs \eqref{eqn:gn} and defining the axial functions $Z_n\left(z_p\right)= \left(n-1\right)\left(n+2\right)\left[\theta_{\rm div}^2/4\right]\left[1+z_p^2/z_R^2\right]^{-1}$ one then finds the complex-valued function $A$ in the forward direction as:
\begin{align}
A \approx&\, \frac{\theta_{\rm max}^2}{4\left[1+z_p^2/z_R^2\right]}\sum_{n=1}^{\infty}\,\left(2n+1\right) \exp\left(-Z_n\right)\left[\cos\left(Z_n\frac{z_p}{z_R}\right)+i\sin\left(Z_n\frac{z_p}{z_R}\right)\right]\approx \frac{\theta_{\rm max}^2}{4\left[1+z_p^2/z_R^2\right]} \left(\frac{4}{\theta_{\rm div}^2} + i\frac{4 z_p^2/z_R^2}{\theta_{\rm div}^2}\right).\label{eq:AppendixA0}
\end{align}
The last approximate equality was achieved by replacing the discrete sums via their corresponding integrals, i.e.\ $\int_1^{\infty}\left(2n+1\right)\exp\left(-Z_n \right)\cos\left(Z_n z_p/z_R\right)\mathrm{d}n$ and $\int_1^{\infty}\left(2n+1\right)\exp\left(-Z_n \right)\sin\left(Z_n z_p/z_R\right)\mathrm{d}n$. This approximation works better for the imaginary part and for paraxial beams with $\theta_{\rm div}\ll 1$, i.e.\ when the integrand is changing mildly with increasing $n$. The real part's approximation can be improved by taking $1/2$ as the lower limit and considering only the constant value at $z=0$, yielding a correction factor of $\exp\left(5\theta_{\rm div}^2/16\right)$. The real and imaginary part of the axial function $A$, eqs.\ \eqref{eq:ARe} and \eqref{eq:AIm} of the main article, then follow from the above expression \eqref{eq:AppendixA0}.

For the largest collection angle $\theta_{\rm max}=\pi$ the sum in eq.\ \eqref{eq:AppendixExtR} vanishes exactly. This may be seen by noting that $\Pi_n\left(\pi\right)=\left(-1\right)^{n+1}\Pi_n\left(0\right)$ and $\tau_n\left(\pi\right)=\left(-1\right)^{n}\Pi_n\left(0\right)$ (B.12 - B.13) whereby the numerator in the fraction becomes zero. Therefore, $A=|g_1|^2$ is purely real-valued. This limiting behaviour already occurs for collection angles $\theta_{\rm max}\gtrsim 2\theta_{\rm div}$, which was verified numerically. Consequently $\sigma_{\rm ext}^\pi=-|g_1|^2 k\mathfrak{I}\left(\alpha\right)\approx \sigma_{\rm abs}^\pi$ for absorbing particles. If the polarizability from the Clausius Mossoti relation $\alpha_{\rm CM}$ is real-valued, then the dipolar Mie coefficient must be expanded up to $a_1=\left[\dots\right] i + k^6 |\alpha_{\rm CM}|^2 /36\pi^2$ to find a non-zero real value and consequently the correct fractional extinction coefficient in eq.\ \eqref{eq:AppendixSigmaExtOnAxis1}. Equivalently, the polarizability may be corrected for radiation back reaction via $\alpha\rightarrow \alpha_{\rm CM}/\left[1+ik^3\alpha_{\rm CM}/6\pi\right]\approx \alpha_{\rm CM} - ik^3|\alpha_{\rm CM}|^2/6\pi$. Then one finds a value of $\sigma_{\rm ext}^\pi=|g_1|^2 k^4|\alpha_{\rm CM}|^2/6\pi \approx |g_1|^2 \sigma_{\rm sca}^R=\sigma_{\rm sca}^\pi$ which the fractional cross-section attains for angles $\theta_{\rm max}\gtrsim 2\theta_{\rm div}$. This also shows that the absorption reads $\sigma_{\rm abs}^\pi=-|g_1|^2 k\mathfrak{I}\left(\alpha_{\rm CM}\right)$ up and including terms of order $x^6$.

\subsection{Small particle limit (off-axis)}
For off-axis scattering we may set $n,n'\rightarrow 1$ in case of the fractional scattering cross-section, Eq.\ \eqref{eq:AppendixSigmaScaOffAxis}, and for the extinction cross-section one may consider $n'=1$ at most in Eq.\ \eqref{eq:AppendixSigmaExtOffAxis}. This is equivalent to considering $m=\left\{-1,0,1\right\}$ only. In this case only $2n_{\rm max}$ integrals must be evaluated once and $3n_{\rm max}$ summands need to be summed up at each position. Also, $b_1 \ll a_1\approx i\alpha k^3/6\pi$ such that only the electric dipolar Mie coefficient is relevant. Then:
\begin{align}
\sigma_{\rm ext}&=k\,\mathfrak{R}\left(\left(i\alpha\right)^{*} \sum_{n}\sum_{m=-1}^{1} \frac{N_n}{2} \left[g_{n,{\rm TM}}^{m}g_{1,{\rm TM}}^{m*} \left(I_1\right)_{n,1}^{|m|}
 + im g_{n,{\rm TE}}^{m}g_{1,{\rm TM}}^{m*} \left(I_2\right)_{n,1}^{|m|}\right]\right)\label{eq:AppendixSigmaExtOffAxisRayleigh}\\
 \sigma_{\rm sca}=&\frac{k^4 |\alpha|^2}{8\pi} \sum_{m=-1}^{1} |g_{1,{\rm TM}}^{m}|^2 \left(I_1\right)_{1,1}^{|m|} .\label{eq:AppendixSigmaScaOffAxisRayleigh}
\end{align}
The fractional extinction may then be written as in Eq.\ \eqref{eq:SigmaExtOffaxisRayleigh} of the article if $(A_x)^{*}$ is identified with the double-sum appearing in Eq.\ \eqref{eq:AppendixSigmaExtOffAxisRayleigh}. For the scattering cross-section only the following two integrals appear: $\left(I_1\right)_{1,1}^0=\left[8-9\cos\left(\theta_{\rm max}\right)+\cos\left(3\theta_{\rm max}\right)\right]/12$ and $\left(I_1\right)_{1,1}^1=\left[16-15\cos\left(\theta_{\rm max}\right)-\cos\left(3\theta_{\rm max}\right)\right]/12$. Noting that the absorption cross-section for small particles reduces to $\sigma_{\rm abs}^\pi \approx -k\mathfrak{I}\left(\alpha\right) [2|g_{1,{\rm TM}}^{-1}|^2+2|g_{1,{\rm TM}}^{+1}|^2+|g_{1,{\rm TM}}^{0}|^2]$, the fractional scattering cross-section is seen to be no longer proportional to the local beam's intensity (the square-bracketed term). Only for $\theta_{\rm max}=\pi$ ($\theta_{\rm max}=\pi/2$) the two integrals reduce to $4/3$ ($2/3$) and $8/3$ ($4/3$), respectively, whereby the proportionality holds again.

\subsection{On the issue of convergence}
The number of multipoles required for convergence of $A$ in eqs.\ \eqref{eqnISI} or \eqref{eq:RAextapprox} also depends on the axial coordinate, requiring more for larger offsets. The BSCs $g_n$ of a focused field are seen to decay with a characteristic multipole order $n_c$. For the Gaussian beam described by the BSCs \eqref{eqn:gn} the critical multipole order is $n_c\approx k \omega_0$ for the particle being in the focus. For larger particle offsets relative to the beam waist the inclusion of even higher multipole orders becomes necessary to adequately account for the incidence field structure, which is in accord with the observation that the magnitude of the BSCs $|g_n\left(z_p\right)|$ decays slower for large $z_p$. Independent on the relative particle position, the situation becomes convergence-wise even worse if near paraxial beams of small divergence are considered. Indeed, even an incident plane wave requires in its spherical base an infinite sum representation with constant BSCs $g_n=1$. The full solid angle integration, however, removes this difficulty and results in the familiar expression of the total extinction cross-section $\sigma_{\rm ext}$. For the evaluation of the total cross-sections only a single term is required to be accurate to within less than a percent of the exact value for small particles with $x\ll 1$. This is a consequence of the orthonormality of the angular basis functions $\Pi_n+\tau_n$ over the full polar angular domain. Only then, the usual statement \cite{BohrenHuffman} can be made that the maximum multipole order necessary is about $n_{\rm max}=2+x+4.3x^{1/3}$. The observation of the bad convergence of the extinction integral when the domain of integration is not the entire solid angle is consistent with the detailed study by M.\ Berg \cite{Berg2008} for plane wave scattering.

\section*{Acknowledgments}
Financial support by the DFG research unit 877 and the graduate school BuildMoNa as well as funding  by the European Union and the Free State of Saxony is acknowledged.

\bibliographystyle{unsrt}

\end{document}